 \documentclass[pra,amsmath,amssymb,twocolumn, showpacs, superscriptaddress,10pt]{revtex4-1}

\usepackage{amsmath}
\usepackage{hyperref}
\usepackage{graphicx}
\usepackage{amsfonts}
\usepackage{amsthm}
\usepackage{cases}
\usepackage{bm}

\usepackage{color}
\definecolor{Blue}{rgb}{0.00, 0.00, 1.00}
\definecolor{Red}{rgb}{1.00, 0.00, 0.00}

\hypersetup{
    colorlinks=true,       
    linkcolor=red,          
    citecolor=blue,        
    filecolor=magenta,      
    urlcolor=cyan           
}

\newcommand{\nn}{\nonumber}
\newcommand{\be}{\begin{equation}}
\newcommand{\ee}{\end{equation}}
\newcommand{\bea}{\begin{eqnarray}}
\newcommand{\eea}{\end{eqnarray}}

\newcommand{\beq}{\begin{equation}}
\newcommand{\eeq}{\end{equation}}
\newcommand{\beqn}{\begin{eqnarray}}
\newcommand{\eeqn}{\end{eqnarray}}

\DeclareMathOperator{\Ai}{Ai}

\newcommand{\x}{{\bf x}}
\newcommand{\y}{{\bf y}}
\newcommand{\p}{{\bf p}}

\begin{document}

\title{Wigner function of noninteracting trapped fermions}

\author{David S. \surname{Dean}}
\affiliation{Univ. Bordeaux and CNRS, Laboratoire Ondes et Mati\`ere  d'Aquitaine
(LOMA), UMR 5798, F-33400 Talence, France}
\author{Pierre Le Doussal}
\affiliation{CNRS-Laboratoire de Physique Th\'eorique de l'Ecole Normale Sup\'erieure, 24 rue Lhomond, 75231 Paris Cedex, France}
\author{Satya N. \surname{Majumdar}}
\affiliation{LPTMS, CNRS, Univ. Paris-Sud, Universit\'e Paris-Saclay, 91405 Orsay, France}
\author{Gr\'egory \surname{Schehr}}
\affiliation{LPTMS, CNRS, Univ. Paris-Sud, Universit\'e Paris-Saclay, 91405 Orsay, France}

\date{\today}

\begin{abstract}
We study analytically the Wigner function $W_N(\x,\p)$ of $N$ noninteracting fermions trapped in a smooth confining potential $V(\x)$ in $d$ dimensions. At zero temperature, $W_N(\x,\p)$ is constant over a finite support in the phase space $(\x,\p)$ and vanishes outside. Near the edge of this support, we find a universal scaling behavior of $W_N(\x,\p)$ for large $N$. The associated scaling function is independent of the precise shape of the potential as well as the spatial dimension $d$. We further generalize our results to finite temperature $T>0$. We show that there exists a low temperature regime $T \sim e_N/b$ where $e_N$ is an energy scale that depends on $N$ and the confining potential $V(\x)$, where the Wigner function at the edge again takes a universal scaling form with a $b$-dependent scaling function. This temperature dependent scaling function is also independent of the potential as well as the dimension $d$.  Our results generalize to any $d\geq 1$ and $T \geq 0$ the $d=1$ and $T=0$ results obtained by Bettelheim and Wiegman [Phys. Rev. B {\bf 84}, 085102 (2011)].

\end{abstract}



\maketitle

\section{Introduction}

\subsection{Background}

The Heisenberg uncertainty principle, the basic cornerstone of quantum mechanics, tells us that the position and the momentum
of a single quantum particle cannot be measured simultaneously. In position space, the squared wave function $|\psi(x)|^2$ 
is the probability density. Similarly, $|\hat \psi(p)|^2$ (where $\hat \psi(p)$ is the Fourier transform of $\psi(x)$)
gives the probability density in momentum space. Although the joint probability density function (PDF)
cannot be defined in phase space $(x,p)$, the closest object to such a joint PDF is the celebrated single particle
``Wigner function'' \cite{wigner}
\be \label{wigner1}
W_1(x,p) =  \frac{1}{2\pi \hbar} \int_{-\infty}^{+\infty} dy \, e^{i py/\hbar} \psi^*(x + \frac{y}{2}) \psi(x - \frac{y}{2})  \;.
\ee 
By integrating $W_1$ over $p$ one recovers the spatial PDF, $|\psi(x)|^2$, and similarly, by 
integrating $W_1$ over $x$ one recovers the momentum PDF, $|\hat \psi(p)|^2$. However in general
$W_1(x,p)$ need not be positive, and hence does not have the interpretation of a joint PDF. 
Nevertheless the Wigner function has been  useful in numerous contexts \cite{case,Bazarov}, including 
quantum chaos and semiclassical physics \cite{berry1,Hannay}, in quantum optics \cite{bookQuantumOptics},
in the modeling of optical devices \cite{Bazarov}, and in quantum information \cite{measurementWigner}.
The Wigner function has been measured experimentally in various contexts, for instance in   quantum state tomography \cite{MeasurementWigner} and in trapped atom set-ups \cite{Atom experiment}. 
It has also been used in many body systems, e.g. in 
Bose-Enstein condensates \cite{Impens}, the implementation of  numerical 
methods for fermions \cite{FermionsWignerMC} and more recently in the
context of non-equilibrium dynamics of a perturbed Fermi gas \cite{Wiegman}. 

Recently there has been considerable interest in trapped Fermi gases, both theoretically~\cite{GPS08} and in cold atom experiments~\cite{BDZ08}. 
Even in the non-interacting limit this system displays rich and universal quantum and thermal fluctuations, as was demonstrated
recently \cite{Kohn,Eis2013,us_finiteT,DPMS:2015,fermions_review,CMV2011,marino_prl,farthest_f}. The case of the harmonic trap played a fundamental role
because it is solvable and makes an important connection between trapped non interacting fermions and the eigenvalues of a random matrix. Indeed
in one-dimension ($d=1$) and at zero temperature $T=0$, the positions of the fermions are in one-to-one correspondence with the eigenvalues of the Gaussian Unitary Ensemble (GUE) of Random Matrix Theory (RMT)~\cite{CMV2011,Eis2013,marino_prl,fermions_review}. Consequently, at $T=0$, the quantum fluctuations of 
$N$ fermions, characterized by the squared many body ground state wave function, $|\Psi_0(x_1, \ldots, x_N)|^2$, was shown to
be identical to the joint PDF of the eigenvalues of a GUE random matrix. Similarly, the joint distribution of the momenta is given by
$|\hat \Psi_0(p_1, \ldots, p_N)|^2$, where $\hat \Psi_0$ is the $N$-variable Fourier transform of $\Psi_0$.
In the case of the harmonic trap, because of the
symmetry $x \leftrightarrow p$ (in scaled units), it is identical to the joint PDF of the positions,
i.e. $\hat \Psi_0=\Psi_0$. 

The squared many-body wave function $|\Psi_0(x_1, \ldots, x_N)|^2$ (in real space) or its Fourier counterpart $|\hat \Psi_0(p_1, \ldots, p_N)|^2$ (in momentum space) encodes  information about quantum fluctuations. For instance, by   
integrating $|\Psi_0(x_1, \ldots, x_N)|^2$ (respectively $|\hat \Psi_0(p_1, \ldots, p_N)|^2$) over $N-1$ positions (respectively momenta), one obtains the average
density of fermions in real space (respectively in momentum space). In the $N \to +\infty$ limit, from the mapping to the GUE, it is 
given (in scaled units and normalized to unity) by the Wigner semi-circle law of RMT, $\rho_W(y) = \pi^{-1} \sqrt{2-y^2}$, with $|y| \leq \sqrt{2}$ 
and zero elsewhere. Near the soft edge $y=\sqrt{2}$ (and similarly around $y = -\sqrt{2}$), the density gets smeared over a width $w_N \sim N^{-1/6}$ 
which defines the edge regime, and the density profile is described by a non trivial scaling function $F_1$,
known in RMT \cite{BB91,For93}. These results extend to all $n$-point correlation functions, either in position or momentum space.
In particular the scaled PDF of the position of the rightmost fermion, $x_{\max}=\max_{1\leq i\leq N} x_i$, 
is
given \cite{us_finiteT,fermions_review} by the Tracy-Widom (TW) distribution of the GUE \cite{TW}.
Interestingly, the (scaled) largest fermion momentum $p_{\max}=\max_{1\leq i\leq N} p_i$, measurable
in time of flight experiments \cite{TOF}, is also distributed with the same TW distribution.
This analysis has been recently extended to any spatial dimension $d$ \cite{DPMS:2015,farthest_f}, to 
finite temperature and beyond the harmonic oscillator for more general smooth potentials
\cite{fermions_review}.

It is thus natural to ask which of these universal properties extend to the Wigner function for $N$ noninteracting trapped fermions, to gain insight on the quantum fluctuations in the
phase space. The $N$ body Hamiltonian is $\hat {\cal H}_N=\sum_{i=1}^N \hat H(\hat\x_i,\hat\p_i)$, where the 
single particle Hamiltonian for spinless fermions of mass $m$ is given by
\be
\hat H = \hat H(\hat\x,\hat\p)=\frac{\hat \p^2}{2 m} + V(\hat \x) ,\label{ham} 
\ee
with $V(\x)=\frac{1}{2} m \omega^2 \x^2$ in the case of the harmonic oscillator. 
The many body Wigner function is defined at $T=0$ as a generalization for any $N$ and $d$ of \eqref{wigner1} 
\bea\label{def_W}
W_N({\bf x},\p) & = &\frac{N}{(2\pi \hbar)^d} \int_{-\infty}^{+\infty} d{\bf y} \, d\x_2 \ldots d\x_N  \,
e^{\frac{i \p \cdot {\bf y}}{\hbar}} \\
& \times& \Psi_0^*(\x+\frac{{\bf y}}{2}, \x_2,\ldots, \x_N)  \Psi_0(\x-\frac{{\bf y}}{2}, \x_2,\ldots, \x_N) ,\nn \label{wig1def}
\eea
which by construction satisfies
\bea
&& \!  \int_{-\infty}^{+\infty} d\p \, W_N({\bf x},\p)= \rho_N(\x), \nn \\
&&   \int_{-\infty}^{+\infty} d\x \, W_N({\bf x},\p)= \bar \rho_N(\p), \nn \\
&&   \int_{-\infty}^{+\infty} d\x \, d\p \, W_N({\bf x},\p)=N, \label{norm1} 
\eea
where $\rho_N(\x)$ is the average density of fermions (here normalized to $N$), and $\bar \rho_N(\p)$ its counterpart
in momentum space. 

\subsection{Main results}

In this paper, we compute $W_N(\x,\p)$ exactly in the large $N$ limit, both in the bulk
and at the edge of a noninteracting Fermi gas trapped by a confining potential $V(\x)$. We perform the derivation in arbitrary dimension $d$, first at $T=0$ and for the harmonic oscillator, and then at finite temperature and for a large
class of smooth potentials. Our results generalize the result obtained by Bettelheim and Wiegmann \cite{Wiegman} in $d=1$ and at $T=0$. 

\vspace*{0.3cm}

\begin{figure}
\includegraphics[width = \linewidth]{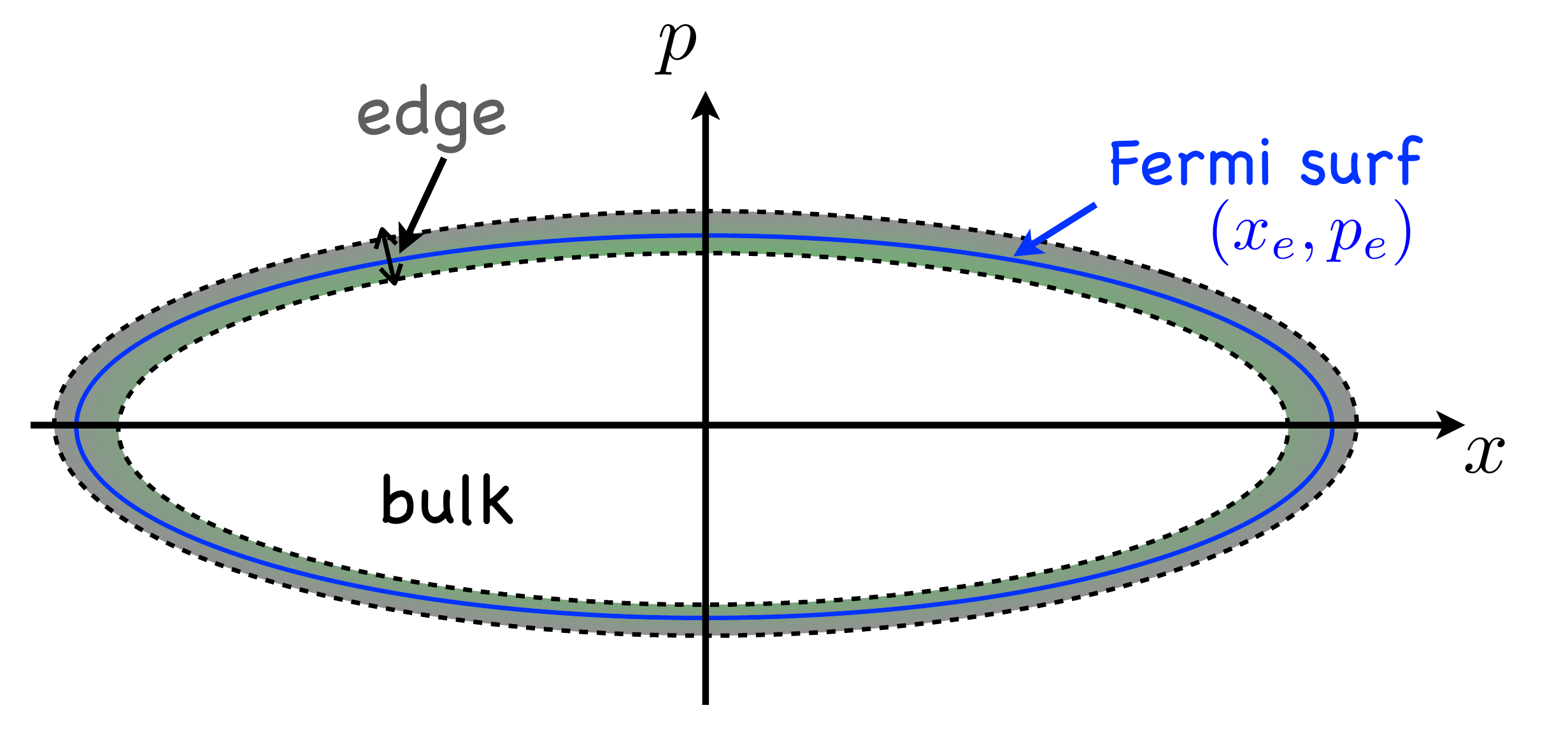}
\caption{Schematic representation of the Fermi surf $(x_e,p_e)$ (blue solid line) defined by Eq. (\ref{def_surf}) in the phase space $(x,p)$. The grey shaded region represents the edge around the Fermi surf, while the white inner region represents the bulk.}\label{Fig_fermi_surf}
\end{figure}

\noindent {\bf Zero temperature $T=0$}:
The result in the bulk is particularly 
simple 
\be
W_N({\bf x},\p)  \simeq \frac{1}{(2 \pi \hbar)^d} \Theta( \mu - E({\bf x},\p) ),  \label{W0bulk}
\ee
where 
\be 
E({\bf x},\p) =
 \frac{\p^2}{2 m} + V({\bf x}),  \label{classicalE}
 \ee
is the classical energy in the phase space. 
Here $\Theta(x)$ is the Heaviside unit step function and $\mu$ is the Fermi energy 
which is related to $N$ via the normalization \eqref{norm1}.
Note that Eq. \eqref{W0bulk}
is valid for large $N$ (equivalently large $\mu$) and for an arbitrary potential $V(\bf x)$. This result, which can be obtained by semi-classical methods
such as the local density approximation (see e.g. \cite{castin}), is obtained here
through a controlled asymptotic analysis of an exact formula. Clearly, the form of $W_N(\x,\p)$, given in Eq. (\ref{W0bulk}), vanishes beyond the surface parametrized by $(\x_e,\p_e)$ where
\beq\label{def_surf}
\frac{\p_e^2}{2m} + V(\x_e) = \mu \;.
\eeq 
Following Ref. \cite{Wiegman}, we will call this surface the ``Fermi surf'', it is the semi-classical version of the Fermi surface in classical phase space (see Fig. \ref{Fig_fermi_surf}). 

Note that by integrating \eqref{W0bulk} over momentum as in 
\eqref{norm1}, one recovers the result for the average number density $\rho_N(\x)= (2\hbar)^{-d} [\p_e(\x)]^d/\gamma_d$
with $\gamma_d= \pi^{d/2} \Gamma(1+d/2)$, where $\p_e(\x):=\sqrt{ 2 m (\mu - V(\x))}$ is
the Fermi momentum, which sets the typical inverse inter-particle spacing $\propto~\hbar(|\p_e(\x)|)^{-1}$. A similar result can be obtained for the average momentum density $\bar \rho_N(\p)$ by integrating \eqref{W0bulk} over $\x$. Both the position as well as the momentum densities exhibit {\em marginal} edges $\x_{em}$ (respectively $\p_{em}$)
beyond which they vanish. The number density $\rho_N(\x)$ vanishes at $\x = \x_{em}$ where $\x_{em}$ satisfies $V(\x_{em}) = \mu$. Similarly, the average momentum density $\bar \rho_N(\p)$ vanishes at $\p = \p_{em}$, where one can show that $|\p_{em}|=\max_{\x} |\p_{em}(\x)| = \sqrt{\max_{\x} \left[{2 m (\mu - V(\x))}\right]}$.
In the case of
the harmonic oscillator, $|\x_{em}|=r_e=\sqrt{2 \mu/m \omega^2}$ and $\p_{em} = \sqrt{2 m\, \mu}$.

\begin{figure}
\includegraphics[width = 0.9\linewidth]{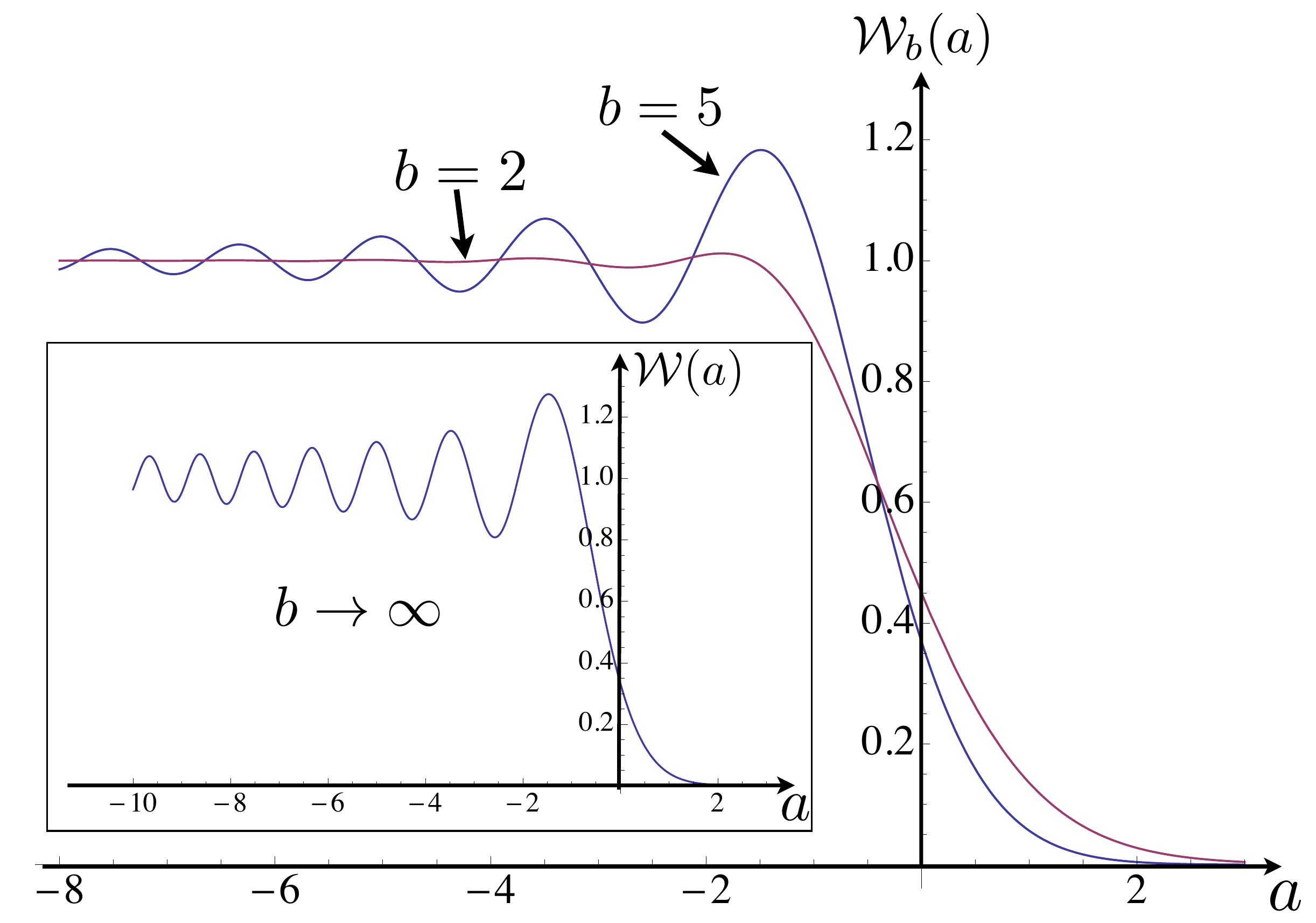}
\caption{Plot of the universal scaling function ${\cal W}_b(a)$, given in Eq. (\ref{asympt_Wb1_intro}), for two different values of the scaled inverse temperature $b = 2$ and $b=5$. In the inset, we show a plot of the zero temperature (i.e. $b \to \infty$) scaling function ${\cal W}_\infty(a) \equiv {\cal W}(a)$ given in Eq. (\ref{scal2}). The oscillations become more pronounced as $b$ increases.}\label{Fig:scaling_function}
\end{figure}

In this paper, our main results concern the properties of $W_N(\x,\p)$ near the Fermi surf in the
$(\x,\p)$ plane, both at $T=0$ and $T>0$, in arbitrary dimensions $d$ and for smooth confining potentials $V(\x) \sim |\x|^p$ for large $|{\bf x}|$. Let us first state our results for $T=0$. In this case, we first define a dimensionless variable $a$ 
\be
a = \frac{1}{e_N} (E({\bf x},{\bf p}) -  \mu),  \label{ae} 
\ee
where $(\x, \p)$ is a point in the phase space close to the Fermi surf and $e_N$ is an energy scale given by
\beq
e_N = \frac{(\hbar)^{2/3}}{(2 m)^{1/3}} \left( \frac{1}{m} (\p_e \cdot  \nabla)^2 V({\bf x}_e)  + |\nabla V({\bf x}_e)|^2 \right)^{1/3} \;. \label{eN_intro} 
\eeq 
We then show that the Wigner function $W_N(\x,\p)$, at $T=0$ and in arbitrary $d$, can be expressed as a universal function of the dimensionless variable $a$ as 
\be
W_N({\bf x},\p)  \simeq \frac{{\cal W}(a)}{(2 \pi \hbar)^d}  \,,\label{scal1}
\ee
where the scaling function
\be
{\cal W}(a) = \int_{2^{2/3} a}^{+\infty} \Ai(u) du \label{scal2} 
\ee
is independent of the space dimension $d$. In Eq. (\ref{scal2}), ${\Ai}(u)$ is the Airy function. The function ${\cal W}(a)$ has the asymptotic behaviors
\begin{eqnarray}
{\cal W}(a) \sim
\begin{cases}\label{asymp_plus}
&(8\pi)^{-1/2}\, a^{-3/4}\,
\exp\left[-\frac{4}{3}\,
a^{3/2}\right]\;, \; a\to + \infty  \\
&\\
& 1\;, \;  \hspace*{4.3cm} a\to -\infty \,. 
\end{cases}
\end{eqnarray}
In particular, the limit $\lim_{a \to -\infty} {\cal W}(a)=1$ ensures
a smooth matching with the bulk result \eqref{W0bulk}. In the inset of Fig. \ref{Fig:scaling_function} we show a plot of this function ${\cal W}(a)$. Note that in $d=1$ our results coincide exactly with the one obtained by Bettelheim and Wiegmann who used 
a completely different method, using a semi-classical analysis of coherent states~\cite{Wiegman}. Our results here provide a generalization of the $d=1$ result to arbitrary $d$. Note that for the case of harmonic oscillator, where $V(\x) = (1/2) m\, \omega^2 \x^2$, the energy scale $e_N$ in \eqref{eN_intro} reduces to
\be
e_N = m \omega^2\,r_e \, w_N \label{eNH}
\ee
where $r_e = \sqrt{2\mu/(m \omega^2)}$ and 
\be \label{wN1} 
w_N = \frac{1}{\alpha \sqrt{2}} (\mu/\hbar \omega)^{-1/6} \;, \; {\rm with}\; \alpha=\sqrt{m\omega/\hbar} \;, \;
\ee
represents the width of the edge region in the real space \cite{DPMS:2015,fermions_review}. Hence the energy scale $e_N$ for the harmonic oscillator reads
\beq\label{eN_OH}
e_N =  (\hbar \omega)^{2/3}\, \mu^{1/3} \;.
\eeq
Furthermore, for the harmonic oscillator, the Fermi energy $\mu$ is related to $N$, for large $N$, via~\cite{DPMS:2015,fermions_review}
\be\label{Fermi_energy}
\mu \sim \hbar \omega \left(N \, \Gamma(d+1) \right)^{1/d} \;.
\ee
Consequently, the argument $a$ of the scaling function ${\cal W}(a)$ in Eq. (\ref{ae}) reduces, in this case, to
\be
a = \frac{1}{w_N} \left(\sqrt{\frac{\p^2}{m^2 \omega^2} + \x^2}- r_e\right) \label{scal1_OH} \;.
\ee

\vspace*{0.3cm}
\noindent{\bf Finite temperature $T>0$:}  Next, we generalize our $T=0$ results for the Wigner function to finite temperature $T$. As in the $T=0$
case, there are two regimes, namely the bulk and the edge. The sharp bulk behavior at $T=0$ in Eq. (\ref{W0bulk}) is
smeared out by thermal fluctuations at finite $T$ and is replaced by  
\be
W_{\tilde \mu}({\bf x},\p) = \frac{1}{1 + e^{\beta ( \frac{\p^2}{2 m } + V(\bf x) - \tilde \mu)}}, \label{WTbulk_intro} 
\ee 
where $\beta = 1/T$. The finite temperature chemical potential $\tilde \mu$, in the canonical ensemble, 
can be determined as a function of $\beta$ and $N$ via the Fermi relation 
\be
N = \sum_{\bf k} \langle n_{\bf k} \rangle = \sum_{\bf k} \frac{1}{1+ e^{\beta(\epsilon_{\bf k}-\tilde \mu)}}, \label{Nmut_intro}
\ee 
where the $\epsilon_{\bf k}$ denote the single particle energy levels of the Hamiltonian $\hat H$ in \eqref{ham}. In the limit $T\to 0$, $\tilde \mu \to \mu$ 
from Eq. (\ref{Nmut_intro}) and Eq. (\ref{WTbulk_intro}) reduces to the $T=0$ result in (\ref{W0bulk}). This semi-classical finite temperature bulk result in Eq.~(\ref{WTbulk_intro}) was also derived by other methods \cite{bartel1985}.

Near the finite temperature edge, where $E(\x,\p) \to \tilde \mu$, we show that the Wigner function has a universal scaling behavior for large $N$.
This universal behavior emerges when the temperature $T$ scales with $N$ (or equivalently with $\mu$) in a particular fashion, namely when temperature $T \sim e_N$ where $e_N$ is the energy scale defined in Eq. (\ref{eN_intro}). Note that $e_N$ just depends on the Fermi energy $\mu$, but not on the temperature. Hence we set 
\be\label{relation_beta_eN}
\beta \, e_N = b \;,
\ee  
where the dimensionless parameter $b = O(1)$ is kept fixed in the limit of large $N$. For instance, for the harmonic oscillator in $1d$, using Eq. (\ref{eN_OH}) and $\mu \sim \hbar \omega \,N $, one gets  $b  = (\hbar \omega/T) N^{1/3}$. This is the same temperature scale that appears in the analysis of the spatial correlations near the edge in real space \cite{fermions_review}. In this temperature regime, one can show that the finite temperature chemical potential $\tilde \mu \sim \mu$, indicating that the finite temperature edge is the same as the zero temperature edge. Hence, as in the $T=0$ case (\ref{ae}), we consider the same dimensionless variable $a$. We show that in this temperature regime characterized by the single dimensionless parameter $b$ (\ref{relation_beta_eN}), the Wigner function takes a scaling form
\be
W_{\tilde \mu}(\x,\p) \sim \frac{{\cal W}_b(a)}{(2 \pi \hbar)^{d}} \;,\quad   a = 
\frac{1}{e_N} \left(\frac{\p^2}{2 m} + V({\bf x}) -  \mu\right),
\ee
with $e_N$ given in Eq. (\ref{eN_intro}). The scaling function ${\cal W}_b(a)$ is given by
\bea\label{asympt_Wb1_intro} 
 {\cal W}_b(a) = \int_{-\infty}^\infty dy\, \frac{{\rm Ai}(y)}{1 + e^{a\,b}e^{-b\,y 2^{-2/3}}} \;.
\eea
In Fig. \ref{Fig:scaling_function} we show a plot of ${\cal W}_b(a)$ for two different values of $b = 2$ and $b=5$. 

In the $T \to 0$ limit, i.e. $b \to \infty$ limit, the function ${\cal W}_b(a)$ reduces to ${\cal W}(a)$ given in Eq. (\ref{scal2}), i.e. ${\cal W}_\infty(a) \equiv {\cal W}(a)$. The asymptotic behaviors of ${\cal W}_b(a)$ are given by
\begin{eqnarray}
{\cal W}_b(a) \sim
\begin{cases}\label{asymp_Wb}
&e^{b^3/12}\, e^{-a\,b}\,\;,\; \;\hspace*{2.7cm} a\to + \infty  \\
&\\
& 1\;, \;  \hspace*{4.3cm} a\to -\infty \,. 
\end{cases}
\end{eqnarray}
Note that for any finite $b$, the right tail of ${\cal W}_b(a)$, as $a \to \infty$, decays exponentially with $a$. It is only exactly at $T=0$, i.e.  when $b \to \infty$ limit, that the right tail decays faster than exponentially as in Eq. (\ref{asymp_plus}). Finally, we note that, in the $T=0$ case, this edge scaling function ${\cal W}_b(a)$ is completely universal, i.e., independent of the dimension $d$ as well as the confining potential $V(\x)$ as long as the potential is non-singular.

The rest of the paper is organized as follows. In Section II, we compute the Wigner function at zero temperature. Section II A contains the exact solution (for any finite $N$) for the $1d$ harmonic oscillator, Section II B discusses the $d$-dimensional harmonic oscillator, while in Section II C we generalize these results for arbitrary smooth confining potentials. In Section III, these results are generalized to finite temperature $T>0$. Finally, we conclude with a summary and discussion in Section IV. Some details are relegated to the Appendices.

\section{Wigner function at zero temperature}

At $T=0$, the quantum correlation functions  of noninteracting fermions can be 
written as determinants constructed from a central object, the 
so-called kernel (see e.g. \cite{fermions_review})
\be
K_\mu(\x,\x') = \sum_{\bf k} \Theta(\mu-\epsilon_{\bf k}) \psi_{\bf k}^*(\x) \psi_{\bf k}(\x') \label{Kmu}
\ee
in terms of the single particle eigenfunctions $\psi_{\bf k}(\x)$ of \eqref{ham}
and their associated eigenenergies $\epsilon_{\bf k}$, labeled by quantum numbers ${\bf k}$. 
In \eqref{Kmu} $\mu$ is chosen so that the sum contains exactly $N$ levels. 
It turns out that one can relate the Wigner function \eqref{def_W} to the kernel, using the 
following formula 
\bea
K_\mu(\x,\x') &&= N \int_{-\infty}^{+\infty} d\x_2 \ldots d\x_N  \nn \\
 &\times& \Psi_0^*(\x, \x_2,\ldots, \x_N)  \Psi_0(\x', \x_2, \ldots, \x_N), \label{Kmu2} 
\eea
which follows from the property that $\Psi_0$ is a Slater determinant constructed
from the $\psi_{\bf k}$'s (see the derivation in Appendix \ref{app:A}). Comparing \eqref{Kmu2} and
\eqref{def_W} we obtain 
\be
\! W_N({\bf x},\p)  = \frac{1}{(2\pi \hbar)^d} \int_{-\infty}^{+\infty} d\y \,e^{\frac{i \p \cdot {\bf y}}{\hbar}}\,
K_\mu(\x+ \frac{\y}{2} ,\x - \frac{\y}{2}) .\label{WK} 
\ee
The scaling behavior of this kernel $K_\mu$ has been well studied in the large $\mu$ and $N$ limit \cite{fermions_review}.
One can then use these results in \eqref{WK} to obtain information
about the Wigner function as shown below.

\subsection{Calculation in $d=1$ for the harmonic oscillator}

Let us first present an exact calculation for the $d=1$ harmonic oscillator, using 
space, momentum, time and energy dimensionless units, i.e. in units of 
\be \label{units}
x_0=1/\alpha \quad , \quad p_0=\hbar \alpha \quad , \quad t_0=1/\omega \quad , \quad e_0=\hbar \omega.
\ee
In these scaled units $\mu \simeq N$ for large $N$. The kernel
reads
\be
K_\mu(x,x')= \sum_{k=0}^{N-1} \psi_k(x) \psi_k(x'), \label{K0} 
\ee 
where $\psi_k(x)=(\frac{1}{\sqrt{\pi}2^k k!})^{\frac{1}{2}} H_k(x) e^{-\frac{1}{2} x^2}$
and $H_k$ is the $k$-th Hermite polynomial. Plugging it in Eq. (\ref{WK}), and specifying $d=1$, we
obtain in dimensionless units and at $T=0$
\be\label{wigner_ho.1}
W_N(x,p) = \frac{1}{2\pi} \sum_{k=0}^{N-1} \int_{-\infty}^\infty dy\, e^{i p y} \psi_k(x+ \frac{y}{2}) \psi_k(x- \frac{y}{2}) \; \;.
\ee
Remarkably, the integral over $y$ in Eq. (\ref{wigner_ho.1})
can be performed explicitly using an identity first derived by Groenewold~\cite{Groenewold46}
\bea \label{identity_laguerre}
&& \int_{-\infty}^{+\infty}  dy e^{i p y} \psi_k(x+ \frac{y}{2}) \psi_k(x- \frac{y}{2})  \\
&& = 2 (-1)^k L_k(2 (x^2 + p^2)) e^{- x^2 - p^2} ,\nonumber 
\eea
where $L_k(y)$ is the Laguerre polynomial of degree $k$, defined via its generating function
\begin{equation}
\sum_{k=0}^{\infty} z^k\, L_k(y) = \frac{1}{(1-z)}\, e^{- z\,y/(1-z)}\,.
\label{laguerre_gf}
\end{equation}
Substituting (\ref{identity_laguerre}) in Eq. (\ref{wigner_ho.1}) gives the explicit result
\begin{equation}
W_N(x,p)= \frac{1}{\pi}\, \sum_{k=0}^{N-1} (-1)^k\, L_k\left(2(x^2+p^2)\right)\, e^{-x^2-p^2}\,.
\label{wigner_T0.1}
\end{equation}\\
\begin{figure}
\includegraphics[width = \linewidth]{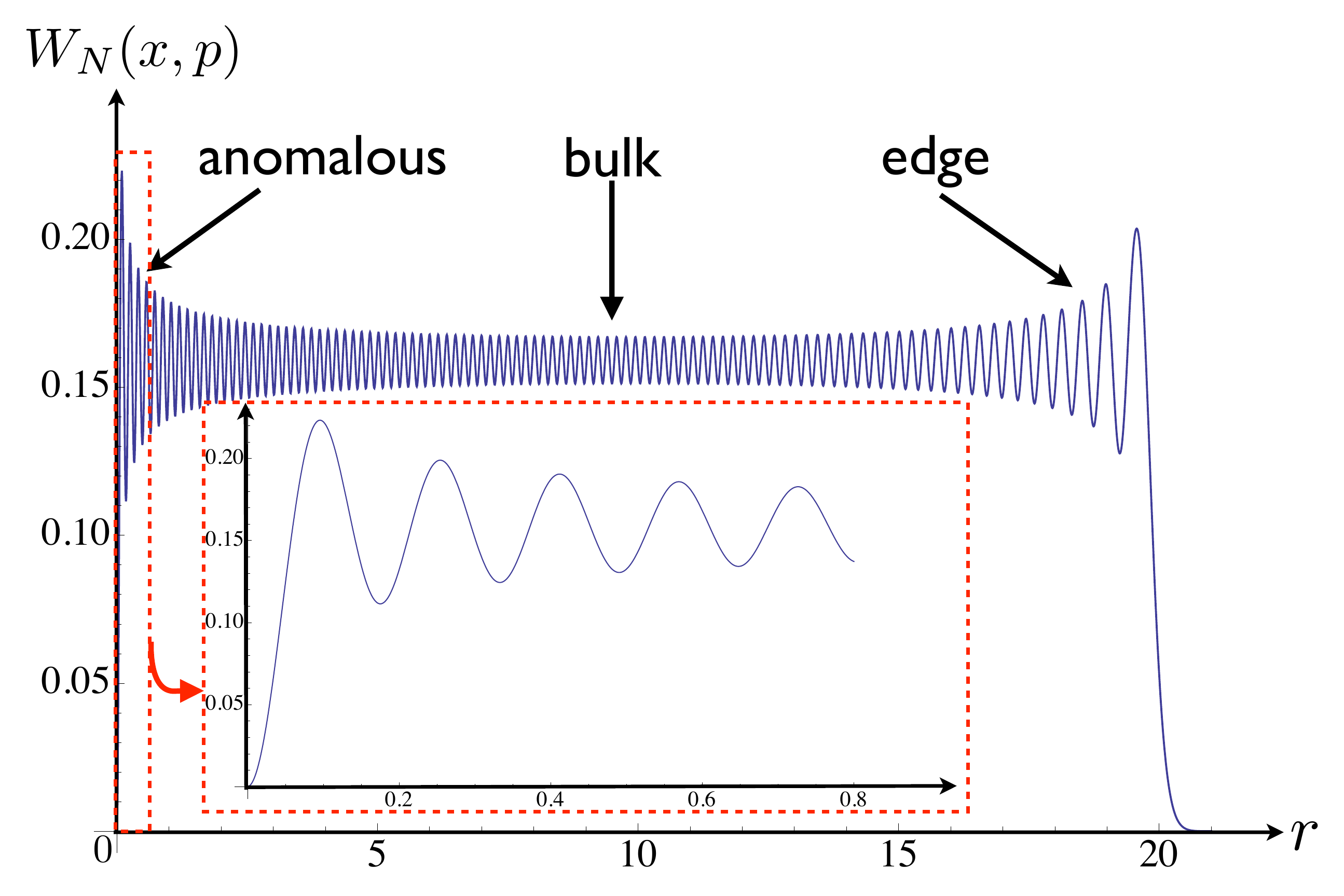}
\caption{Plot of $W_N(x,p)$ as a function of $r = \sqrt{x^2+p^2}$ for the $1d$ harmonic oscillator, as given in Eq. (\ref{wigner_T0.1}) for $N=  200$. {\bf Inset}: Zoom on the range $r \in [0,0.8]$. Note in particular that $W_N(0,0)=0$ here as $N=200$ is even (see Eq. (\ref{vanish})).} \label{fig:Wigner}
\end{figure}
We have plotted $W_N(x,p)$ as a function of $r=\sqrt{x^2+p^2}$ for  $N=200$
in Fig. \ref{fig:Wigner}. From this figure one sees that there are three distinct regimes
at large $N$: (i) the bulk regime where $W_N(x,p)$ oscillates around the bulk value 
${1}/{(2 \pi)} = 0.159155 \ldots$, (ii) the edge regime around $r = \sqrt{2 N}$ where the Wigner function
vanishes over a width of order $N^{-1/6}$ (see below) (iii) an "anomalous" regime 
near $r=0$. This anomalous regime has been pointed out \cite{berry1} for the single particle case ($N=1$), here we show that it persists for multi-particle systems. At $r=0$ the Wigner function vanishes exactly for $N$ even, and equals $1/\pi$ for
$N$ odd. At small $r \sim 1/\sqrt{N}$ there is a scaling regime describing the
Wigner function near $r=0$. We now study these three regimes in detail.

We start by multiplying Eq. \eqref{wigner_T0.1} by $z^N$ and sum over $N$. We obtain
\be\label{eq_laguerre1}
\sum_{N=1}^\infty z^N W_N(x,p) = \frac{1}{\pi} \sum_{N=1}^\infty z^N \sum_{k=0}^{N-1} (-1)^{k} L_k(2r^2) e^{-r^2},
\ee
with $r^2 = x^2 + p^2$. To perform this double sum, we write $z^N = z^{N-k} \,z^k$ and perform the sums separately over $k$ and $m=N-k$. This gives
\be \label{eq_laguerre2}
\sum_{N=1}^\infty z^N W_N(x,p) = \frac{1}{\pi} \sum_{m=1}^\infty z^m \sum_{k=0}^{\infty} (-z)^{k} L_k(2r^2) e^{-r^2} \;.
\ee
Using the generating function of Laguerre polynomials in Eq. (\ref{laguerre_gf}), we get
\be \label{exactGF}
\sum_{N=1}^\infty z^N W_N(x,p) = \frac{1}{\pi} \frac{z}{1-z^2} e^{-\frac{1-z}{1+z} (x^2+p^2)} \;.
\ee
For convenience, we use $z = e^{-s}$ and get
\be \label{exactGF2}
\tilde W(x,p;s):= \sum_{N=1}^{\infty} W_N(x,p) e^{-N s} = \frac{e^{-s- (x^2+p^2) \tanh \frac{s}{2} }}{\pi (1-e^{-2 s})} \;.
\ee

\medskip

{\bf Bulk behavior:} 
To extract the bulk result we consider the limit $s\ll 1$ with $s (x^2+p^2)$ fixed, leading to
\be
\tilde W(x,p;s) \simeq \frac{e^{-\frac{s}{2} (x^2+p^2)}}{2 \pi s}, 
\ee
which yields, after Laplace inversion
\be
W_N(x,p) \simeq \frac{1}{2 \pi} \Theta(2 N - (x^2+p^2)), \label{W2bulk} 
\ee 
in agreement with the 
general result \eqref{W0bulk}.  

\medskip

{\bf Edge behavior:}
We now show that the exact formula \eqref{exactGF} can be used to derive the edge scaling function
\eqref{scal1}, \eqref{scal2} in $d=1$ for the harmonic oscillator.
Our starting point is the exact relation \eqref{exactGF}.
Inverting the generating function using Cauchy's inversion formula, 
\bea
&& W_N(x,p) \\
&& =  \frac{1}{\pi} \int_{c-i\times \infty}^{c+i\times \infty} \frac{ds}{2 i \pi} \frac{e^{s\,(N-1)}}{\left(1-e^{-2s}\right)}\,
\exp\left[- (x^2+p^2) \, \tanh \frac{s}{2} \right]\, ,\nn
\label{laplace.1}
\eea
where $c$ is to the right of all singularities in the complex $s$-plane.
For large $N$, the most important contributions come from the vicinity of $s\to 0$. We set, near the edge $\sqrt{x^2+p^2}=\sqrt{2N}$,
\begin{equation}
\sqrt{x^2+p^2}= \sqrt{2N} + \frac{1}{\sqrt{2}}\, N^{-1/6}\, a\, ,
\label{edge.1}
\end{equation}
where $a$ denotes the distance (on the scale of $N^{-1/6}$) from the edge, see
\eqref{ae} in $d=1$ and dimensionless units. Using $\tanh(s/2)= s/2- s^3/24$ as $s\to 0$,
we find that the integral in Eq. (\ref{laplace.1}), reduces for large $N$ to
\begin{equation}
W_N(x,p) \approx \frac{1}{2\pi}\, \int_{c-i\times \infty}^{c+i\times \infty} \frac{ds}{2 i \pi s}\, 
\exp\left[- s\, N^{1/3}\, a + s^3\, \frac{N}{12} \right].
\label{laplace.2}
\end{equation}
Rescaling further by setting $t= s\, N^{1/3}$, we finally get, near the edge
\bea\label{edge_scaling.1}
&& W_N(x,p) \approx \frac{1}{2\,\pi}\, {\cal W}(a) ,
\eea
where
\bea\label{def_a_1d}
&& a=\sqrt{2}\, N^{1/6}\, \left(\sqrt{x^2+p^2}-\sqrt{2N}\right) , 
\label{edge_scaling.1}
\eea
and the scaling function ${\cal W}(a)$ is given exactly by
\begin{equation}
{\cal W}(a)= \int_{c-i\times \infty}^{c+i\times \infty} \frac{dt}{2 i \pi t}\,  \exp\left[-t\, a+ \frac{t^3}{12}\right]\, .
\label{scaling_function.1}
\end{equation}
Using the integral representation of the Airy function
\be
{\rm Ai}(x) = \int_{c-i\times \infty}^{c+i\times \infty} \frac{d \tau}{2 i \pi}\,  \exp\left[-\tau\, x+ \frac{\tau^3}{3}\right],\
\ee
we find that this scaling function ${\cal W}(a)$ is indeed given by formula \eqref{scal2}. One can easily check that it has
the asymptotic behaviors given in Eq. (\ref{asymp_plus}). \\
%

{\bf Anomalous behavior near $r=\sqrt{x^2+p^2}=0$}.
It is easy to see that the Wigner function vanishes at $r=0$ for $N$ even.
From the definition \eqref{wigner_ho.1}
\bea
W_N(x=0,p=0) &=& \frac{1}{2 \pi}  \int_{-\infty}^{\infty} dy \sum_{k=0}^{N-1} \psi^*_k(\frac{y}{2}) \psi_k(-\frac{y}{2}) \nn \\
& =& \frac{1}{\pi}  \sum_{k=0}^{N-1} (-1)^k = \begin{cases} & \frac{1}{\pi} \quad N \quad  \text{odd} \\
& 0 \quad N \quad \text{even}
\end{cases} \nn \\
&& \label{vanish} 
\eea 
where we used (i) the orthonormality of the single particle wave functions $\psi_k(x)$, and
(ii) the fact that for the harmonic oscillator potential $\psi_k(-x)= (-1)^k \psi_k(x)$. 
Note the property \eqref{vanish} extends to an arbitrary even potential $V(x)=V(-x)$ 
in $d=1$. 

As we show in Appendix \ref{Appendix:anomalous}, near $r=0$ in a regime where $r \sim 1/\sqrt{N}$, the
Wigner function has the following scaling behavior for large $N$ 
\bea\label{FJ0}
&& W_N(x,p) \sim \frac{1}{2\pi} - (-1)^N F(\sqrt{N(x^2+p^2)}) \nn \\
&&  F(z) = \frac{1}{2\pi} J_0( 2\sqrt{2} z) \;,
\eea
where $J_\nu(x)$ is the Bessel function with index $\nu$. Interestingly, this parity dependence in Eq. (\ref{FJ0}) persists even for large $N$. We have verified the scaling behavior in (\ref{FJ0}) by numerically evaluating $W_N(x,p)$ in 
Eq. \eqref{wigner_T0.1}.

\subsection{$d$-dimensional harmonic oscillator}

In dimension $d>1$ it is more convenient to use a different method using the quantum propagator
to calculate the kernel, and in turn the Wigner function via \eqref{WK}. In addition, as we show later 
this method is more versatile as one can treat more general potentials and demonstrate universal properties
of the Wigner function. The method
relies on the following representation of the kernel in $d$ dimensions for arbitrary single particle Hamiltonian
\cite{fermions_review}
\bea
K_\mu(\x,\x') = \int_{C} \frac{dt}{2 \pi i t} e^{\mu t/\hbar} G(\x,\x';t),  \label{KG}
\eea
where $C$ is the Bromwich contour in the complex plane, and $G(\x,\x';t)=\langle \x'|e^{-\hat H t/\hbar}| \bf x\rangle$ is the one particle Euclidean
quantum propagator associated to the Hamiltonian $\hat H$ in \eqref{ham}. Let us apply this relation to the case of the 
harmonic oscillator in dimension $d$ for which the exact propagator is known \cite{Feynman}. We work again here in the aforementioned
dimensionless units in which the propagator reads
\be
G(\x,\x';t) = \frac{1}{(2 \pi \sinh t)^{d/2}} e^{ - \frac{(\x-\x')^2 + (\x^2 + (\x')^2) (\cosh t-1)}{2 \sinh t}}. \label{GHO}
\ee
We first insert  \eqref{GHO} into \eqref{KG} and then use Eq. (\ref{WK}). Performing the
Gaussian integration over $\y$ we obtain
\be
W_N({\bf x},\p)  = \frac{1}{(2\pi)^d} \int_{C} \frac{dt}{2 \pi i t} e^{\mu t}
 \frac{1}{(\cosh \frac{t}{2})^{d}} e^{ - (\x^2+\p^2) \tanh\frac{t}{2}} \;,
\ee 
an exact formula. Note that the $N$-dependence of $W_N(\x, \p)$ is only through $\mu$. As in the $d=1$ case we now analyze this formula both in the bulk as well as in the edge
regime.

{\bf Bulk behavior:} in the bulk one can show that the values of $t$ which dominate the integral
are $O(1/\mu)$. Hence we only need to expand the factor $\tanh(t/2)$ inside the exponential
only to $O(t)$. This leads to 
\be \label{exp1}
W_N({\bf x},\p)  = \frac{1}{(2 \pi)^d} \int_{C} \frac{dt}{2\pi i t} 
e^{ (\mu - \frac{\p^2 + \x^2}{2} ) t }  .
\ee 
The integral over $t$ just gives a Heaviside theta function,
hence establishing the result for the bulk in Eq. \eqref{W0bulk}
in the case of the harmonic potential.

{\bf Edge behavior:} to analyze the edge behavior in the large $\mu$ limit, we need to expand the exponential up to order $t^3$, as 
in \cite{fermions_review} for the study of the edge in real space. We
obtain (discarding terms of $O(t^4)$ in the exponential)
\be \label{exp1}
W_N({\bf x},\p)  =  \int_{C} \frac{dt}{(2\pi)^{d+1} i t} 
e^{ (\mu - \frac{\x^2+\p^2}{2}) t  - \frac{d t^2}{8} + (\x^2+\p^2)  \frac{t^3}{24} }  .
\ee 
Keeping only the leading $O(t)$ term in the exponential in \eqref{exp1}, immediately leads to the
result in the bulk \eqref{W0bulk}. Precisely at the edge, the coefficient of the $O(t)$ term
vanishes. Hence, to study the vicinity of the edge one must keep terms up to $O(t^3)$. 
To this aim, we parameterize the distance to the edge as $\sqrt{\x^2+\p^2} - \sqrt{2 \mu} = w_N a$, with $w_N \ll \sqrt{2 \mu}$
to be determined below, and $a=O(1)$. We now expand
the argument of the exponential \eqref{exp1}, denoted by $S$, 
\be \label{SS}
S = - \sqrt{2 \mu} a w_N t  - d \frac{t^2}{8}+ 2 \mu  \frac{t^3}{24} +O(t^4, w_N t^3 \sqrt{\mu}) \;.
\ee
Let us now define $t= t_N \tau$ where $\tau=O(1)$. To determine the width $w_N$ and the
parameter $t_N$ of the edge regime, the only consistent choice is to impose that both
terms $w_N t$ and $\mu t^3$ in \eqref{SS} are $O(1)$. This leads to $t_N=2^{2/3} \mu^{-1/3}$ and 
$w_N=\frac{1}{\sqrt{2}} \mu^{-1/6}$ and to
\bea
S = - a 2^{2/3} \tau  + \frac{\tau^3}{3} + O(\mu^{-2/3}).
\eea
One thus obtains, upon restoring the physical units, the scaling form \eqref{scal1} for the Wigner function at large $\mu$ as
\be
 W_N({\bf x},\p) \simeq \frac{{\cal W}(a)}{(2 \pi)^d} \quad , \quad 
{\cal W}(a) =  \int_{C} \frac{d\tau}{2 \pi i \tau} e^{ - a 2^{2/3} \tau  + \frac{\tau^3}{3} } ,\label{rep} 
\ee
which is precisely the integral representation given in Eq.~(\ref{scaling_function.1}) with the substitution
$2^{2/3}\, \tau = t$. Thus we obtain the remarkable result that the edge scaling form
${\cal W}(a)$ of the Wigner function for the harmonic oscillator is completely 
independent of the space dimension $d$. The same result can also be obtained (see Appendix \ref{appb}), 
directly from the known scaling behavior of the kernel \cite{fermions_review}. A natural question is whether this
scaling form is also universal with respect to the details of the shape of the confining
potential, as we will discuss below.

In $d=1$, we have seen in the previous subsection that there is an additional anomalous regime when $x^2 + p^2 = O(1/N)$. In $d>1$, similar anomalous regimes are likely to exist, though we have not investigated them in detail.

\subsection{Wigner function for other smooth confining potentials: beyond the harmonic oscillator}
\label{sec:V} 

The case of the harmonic oscillator, treated in the previous section, is special because ${\bf x}$ and
${\bf p}$ (appropriately rescaled in the dimensionless units \eqref{units}) play 
symmetric roles, and the Wigner function depends only
on the single variable ${\bf x}^2 + {\bf p}^2$. For a general potential $V(\bf x)$
this is no longer the case, and a different treatment is needed, as we now show.

Consider a more general smooth potential $V(\bf x)$ in $d$ dimensions. In this case we show,
using the above propagator method, that in the bulk the Wigner function is given by
\eqref{W0bulk}. Putting together formula \eqref{KG} and \eqref{WK} one can express the Wigner function
directly in terms of the Euclidean propagator
\bea
&&  W_N({\bf x},\p) \label{Wdirect} \\
&& = \frac{1}{(2\pi \hbar)^d}  
 \int_{C} \frac{dt}{2 \pi i t} e^{\frac{\mu t}{\hbar}} \int_{-\infty}^{+\infty} d\y e^{\frac{i \p \cdot {\bf y}}{\hbar}}\, G\left(\x+ \frac{\y}{2} ,\x - \frac{\y}{2}, t\right) \;. \nn
\eea
Let us recall that the Euclidean quantum propagator satisfies the
Feynman-Kac equation $G({\bf x}, {\bf x'},t)$ 
\be
 \partial_t G = -\hat H G = \left(\frac{\hbar^2}{2 m} \nabla_x^2 - V({\bf x}) \right) G,
\ee 
with $G({\bf x}, {\bf x'},0)=\delta^d({\bf x}- {\bf x'})$. As was shown in \cite{fermions_review},
and further detailed in Appendix \ref{App:short_time}, the large $N$ limit can be obtained from the small
time $t$ expansion of the quantum propagator, which up to $O(t^3)$ reads
\bea
&& G({\bf x}, {\bf x'},t) = \left(\frac{m}{2 \pi \hbar t}\right)^{d/2} \exp[ - \frac{m}{2 \hbar t} ({\bf x}-{\bf x'})^2 ] \label{exp} \\
&& \times \exp\left[ - \frac{t}{\hbar} S_1({\bf x}, {\bf x'}) - 
\frac{t^2}{2 m} S_2({\bf x}, {\bf x'}) + \frac{t^3}{2 m \hbar} S_3({\bf x}, {\bf x'})\right], \nn
\eea 
where $S_1,S_2,S_3$ for an arbitrary potential $V({\bf x})$ are given explicitly
in Eqs. (243-245) of \cite{fermions_review}. 

To analyze the bulk behavior of the Wigner function we only need the leading $O(t)$ term
$S_1$ which then reads
\be
S_1\left({\bf x} - \frac{\bf y}{2}, {\bf x} + \frac{\bf y}{2}\right) = \int_0^1 du \, V\left( {\bf x} + (u - \frac{1}{2}) {\bf y}\right) .
\ee
We substitute this expression of $S_1$ in \eqref{exp} and keep only up to $O(t)$
terms. Next we substitute this propagator in~\eqref{Wdirect} 
\bea
&& W_N({\bf x},\p)  = \frac{1}{(2\pi \hbar)^d}  
 \int_{C} \frac{dt}{2 \pi i t}  \left(\frac{m}{2 \pi \hbar t}\right)^{d/2}  \int_{-\infty}^{+\infty} d\y e^{\frac{i \p \cdot {\bf y}}{\hbar}}\, \nn  \\
 && \times
 \exp\left[ - \frac{m}{2 \hbar t} {\bf y}^2 
 + \frac{t}{\hbar} \left(\mu - \int_0^1 du \, V( {\bf x} + (u - \frac{1}{2}) {\bf y}) \right) \right] \nonumber.\\
\eea
One can show that the values of $t$ which dominate the integral
are $O(1/\mu)$. Hence by rescaling ${\bf y}$ as shown in the
Appendix \ref{App:short_time}, one can neglect the term $(u-1/2) {\bf y}$ in the argument
of $V$, leading, after integration over ${\bf y}$, to
\be \label{exp2}
W_N({\bf x},\p)  = \frac{1}{(2 \pi \hbar)^d} \int_{C} \frac{dt}{2\pi i t} 
e^{ (\mu - \frac{p^2}{2 m} - V({\bf x}) ) t }  .
\ee 
The integral over $t$ just gives a Heaviside theta function,
hence establishing the result for the bulk in Eq. \eqref{W0bulk}.

From Eq. \eqref{W0bulk} it is clear that the Wigner function vanishes 
beyond the boundary of a bounded support, which defines an
edge in phase space, i.e. a surface parameterized by $({\bf x}_e,\p_e)$ which 
satisfy the equation
\be
 \frac{\p_e^2}{2 m} + V({\bf x}_e) = \mu \;. \label{edgeeq}
\ee 
As for the case of the harmonic oscillator, for large but finite $N$ the jump of the Wigner function described by \eqref{W0bulk} 
is smoothed over a scale $w_N$ which now explicitly depends on the potential. However, the appropriately centered and scaled Wigner 
function at any point of the edge surface, is again universal and is given by ${\cal W}(a)$ in \eqref{scal2}. More precisely the Wigner function
takes the following edge scaling form for $\frac{p^2}{2 m} + V(x) -  \mu \sim \mu^{1/3} \ll \mu$
\be
W_N({\bf x},\p)  \simeq \frac{{\cal W}(a)}{(2 \pi \hbar)^d} \;,   \label{scal1V}
\ee
where the scaled variable $a=O(1)$ is now naturally expressed as the ratio of two energies
\be
a = \frac{1}{e_N} \left(\frac{\p^2}{2 m} + V({\bf x}) -  \mu\right) ,\label{aV} 
\ee
where the energy scale $e_N$ is given 
\beq
e_N = \frac{(\hbar)^{2/3}}{(2 m)^{1/3}} \left( \frac{1}{m} (\p_e \cdot  \nabla)^2 V({\bf x}_e)  + |\nabla V({\bf x}_e)|^2 \right)^{1/3} \;. \label{eN} 
\eeq 
For the derivation of these results see Appendix \ref{App:short_time}. For the harmonic oscillator $V({\bf x})= \frac{1}{2} m \omega^2 \x^2$ one finds
\be
e_N =  \hbar \omega \, \left(\frac{\mu}{\hbar \omega}\right)^{1/3} =  m \omega^2 r_e w_N = \frac{\hbar \omega}{2 \alpha^2 w^2_N}
\ee
We thus see that Eq. \eqref{aV} is consistent with Eq. \eqref{ae} and Eq. \eqref{eN} is
consistent with \eqref{eNH}, as discussed earlier. For a general potential the above results are valid on any point $({\bf x}_e,\p_e)$ on the edge surface.
In particular for $\p_e=0$ we have 
\bea
&& a = \frac{1}{e_N} (V({\bf x}) -  \mu) \simeq \frac{1}{e_N} \nabla V({\bf x}_e) \cdot ({\bf x} - {\bf x}_e)  \\
&& e_N = \frac{(\hbar)^{2/3}}{(2 m)^{1/3}} |\nabla V({\bf x}_e)|^{2/3} \label{eN2}
\eea 

Furthermore, to make contact with the $1d$ result derived in Ref. \cite{Wiegman}, let us
consider a point in the $({\bf x},{\bf p})$ plane near the edge $({\bf x}_e, {\bf p}_e)$ where we set ${\bf x} = {\bf x}_{e}$ and ${\bf p} = {\bf p}_e + \tilde{{\bf p}}$ with $|\tilde {\bf p}| \ll 1$ (see Fig. \ref{Fig_fermi_surf}). In particular, if we focus on $d=1$, the formulae (\ref{aV}) and (\ref{eN}) simplify a lot. In this case the formula (\ref{eN}) becomes
\beq\label{eN_1d}
e_N = \frac{(\hbar)^{2/3}}{(2m)^{1/3}} \left(\frac{p_e^2}{m}V''(x_e) + (V'(x_e))^2 \right) \;,
\eeq
and Eq. (\ref{aV}) reduces, to leading order in $\tilde p$, 
\beq\label{a_1d}
a = \frac{ p_e(x_e)}{e_N}\,\tilde p \; \;.
\eeq
Here, $p_e(x_e)$ is the point on the edge surface parametrized by the function
\beq\label{fermi_surf}
p_e^2(x) = 2m(\mu - V(x)) \;.
\eeq
In Ref. \cite{Wiegman}, this $p_e(x)$ was called the ``Fermi surf''. By taking twice the derivative of \eqref{fermi_surf}, the expression for $e_N$ in Eq. (\ref{eN_1d}) simplifies to
\be\label{eN_simple}
e_N = \left(\frac{\hbar^{2}}{2}\right)^{1/3} |p_e''(x_e)|^{1/3}\,\frac{p_e(x_e)}{m} \;.
\ee
Consequently, from Eq. (\ref{a_1d}), one gets
\beq
a = \kappa \, \tilde p \;, \;\;\; \kappa = \left( \frac{\hbar^2}{2} p_e''(x_e)\right)^{-1/3} \label{kappa1} \;.
\eeq
Thus the argument $a$ of the scaling function ${\cal W}(a)$ in Eq. (\ref{scal1V}) reduces, in $d=1$, to
precisely the argument derived in Ref. \cite{Wiegman}.

\section{Wigner function at finite temperature} 

We now extend our analysis to finite temperature.
The Wigner function at temperature $T=1/\beta$ in the canonical ensemble can be defined from the many body density matrix 
$\hat {\cal D}_N = e^{- \beta \hat {\cal H}_N}/Z_N(\beta)$ as
\bea\label{def_WT}
&&W_{N,T}({\bf x},\p) =\frac{N}{(2\pi \hbar)^d} \int_{-\infty}^{+\infty} d{\bf y} \, d\x_2 \ldots d\x_N  \nn,
e^{\frac{i \p \cdot {\bf y}}{\hbar}} \\
&&\times \langle \x+\frac{{\bf y}}{2}, \x_2,\cdots, \x_N| \hat{\cal D}_N|\x-\frac{{\bf y}}{2}, \x_2,\cdots, \x_N \rangle. 
\eea
By decomposing on the basis of eigenvectors $|E \rangle$ of $\hat {\cal H}_N$ one can equivalently
write it as
\bea
W_{N,T}({\bf x},\p) & = & \frac{1}{Z_N(\beta)} \sum_E e^{- \beta E} W_{N,E}({\bf x},\p)  ,\label{WNT} 
\eea
where $Z_N(\beta)=\sum_E e^{- \beta E}$ is the canonical partition sum and
\be
W_{N,E}({\bf x},\p)  = \frac{1}{(2\pi \hbar)^d} \int_{-\infty}^{+\infty} d\y \,e^{\frac{i \p \cdot {\bf y}}{\hbar}}\,
K(\x+ \frac{\y}{2} ,\x - \frac{\y}{2};\{ n_{\bf k} \}), \label{WK3} 
\ee
where the $N$ body eigenstate $|E \rangle$ is labeled by
a set of occupation numbers $n_{\bf k}=0,1$ of the single particle eigenstates,
such that $E= \sum_{\bf k} n_{\bf k} \epsilon_{\bf k}$ and $N=\sum_{\bf k} n_{\bf k}$. The kernel $K(\x ,\x';\{ n_{\bf k} \})$
has the expression
\bea
K(\x ,\x';\{ n_{\bf k} \}) = \sum_{\bf k} n_{\bf k} \psi_{\bf k}^*(\x)  \psi_{\bf k}(\x'), \label{Kn} 
\eea
where $\psi_{\bf k}(\x)$ is the single particle eigenstate labeled by ${\bf k}$. Note that in the $T=0$
limit  \eqref{Kn} reduces to \eqref{K0} and \eqref{WK3} reduces to \eqref{WK}. 

One can also define the Wigner function in the grand canonical ensemble 
with chemical potential $\tilde \mu$ as
\be
W_{\tilde \mu}(\x,\p) = \frac{1}{Z_{gr}(\beta,\tilde \mu)} \sum_N Z_N(\beta) W_{N,T}(\x,\p) e^{\tilde \mu \beta N}, \label{WNmu} 
\ee
where $Z_{gr}(\beta,\tilde \mu) = \sum_N Z_N(\beta) e^{\tilde \mu \beta N}$ is the grand canonical partition function. Substituting 
\eqref{WNT} in \eqref{WNmu} and summing Eq. \eqref{WK3} over the eigenstates $|E \rangle$ an $N$, i.e over the independent
variables $n_{\bf k}$'s, leads to
\be
W_{\tilde \mu}({\bf x},\p)  = \frac{1}{(2\pi \hbar)^d} \int_{-\infty}^{+\infty} d\y \,e^{\frac{i \p \cdot {\bf y}}{\hbar}}\,
\tilde K_{\tilde \mu}\left(\x+ \frac{\y}{2} ,\x - \frac{\y}{2}\right) \label{WK4} 
\ee
where $\tilde K_{\tilde \mu}(\x,\x')$ is the kernel defined in the grand canonical 
ensemble
\bea
 \tilde K_{\tilde \mu}(\x,\x') = \sum_{\bf k} \langle n_{\bf k} \rangle \psi_{\bf k}^*(\x)  \psi_{\bf k}(\x'), \label{Kmut} 
\eea
where 
\be
\langle n_{\bf k} \rangle = \frac{1}{1+ e^{\beta(\epsilon_{\bf k}-\tilde \mu)}} \label{meannk}
\ee 
is the mean occupation
number of state ${\bf k}$, over the Fermi distribution. 

In the grand canonical ensemble one can relate the finite temperature Wigner function
$W_{\tilde \mu}({\bf x}, {\bf p})$, to the zero temperature Wigner function 
$W_{N}({\bf x}, {\bf p})$. To see this, we consider first the zero temperature kernel
in \eqref{Kmu}. Taking a derivative w.r.t. the Fermi energy $\mu$ gives 
\be
\partial_\mu K_\mu(\x,\x') = \sum_{\bf k} \delta(\mu-\epsilon_{\bf k}) \psi_{\bf k}^*(\x) \psi_{\bf k}(\x') .\label{Kmu22}
\ee
Now we start with Eq. \eqref{Kmut} and \eqref{meannk} rewrite them as
\bea
 \tilde K_{\tilde \mu}(\x,\x') &= & \int_{-\infty}^{+\infty} 
\frac{d\mu'}{1+ e^{\beta(\mu'-\tilde \mu)}} \sum_{\bf k}   \delta(\mu' - \epsilon_{\bf k})  \psi_{\bf k}^*(\x)  \psi_{\bf k}(\x') \nn \\
& = &
\int_{-\infty}^{+\infty}  \frac{d\mu'}{1+ e^{\beta(\mu'-\tilde \mu)}} \partial_{\mu'} K_{\mu'}(\x,\x'),
\eea 
where in the last line we used \eqref{Kmu22}. This equation was previously derived as
Eq. (240) in \cite{fermions_review}. We can now apply this formula to the Wigner function
using \eqref{WK4}, leading to 
\be \label{relationW}
W_{\tilde \mu}(\x,\p) = \int_{-\infty}^{+\infty} d\mu' \frac{1}{1+ e^{\beta(\mu'-\tilde \mu)}} \partial_{\mu'} (W_{N}(\x,\p)|_{\mu= \mu'}),
\ee
which is exact in the grand canonical ensemble. We have denoted the integration variable by $\mu'$ to
avoid confusion with the variable $\mu$ which denotes the Fermi energy. Note that
the zero temperature Wigner function $W_{N}(\x,\p)$ depends implicitly on $\mu$. 
This relation \eqref{relationW} will now allow us to derive the
bulk and edge properties of the Wigner function at finite temperature.

The above results are obtained in the grand canonical ensemble where $\tilde \mu$ is 
a given parameter. However our main goal is to describe the Wigner function in the canonical
ensemble where $N$ is fixed. Indeed in the large $N$ limit, adapting the saddle point method
of \cite{fermions_review} to the Wigner function, we expect that 
one can use the grand canonical results for the canonical ensemble,
provided we determine $\tilde \mu$ as a function of $N$ by the relation
\be
N = \sum_{\bf k} \langle n_{\bf k} \rangle = \sum_{\bf k} \frac{1}{1+ e^{\beta(\epsilon_{\bf k}-\tilde \mu)}} .\label{Nmut}
\ee 
Note that by definition of $\mu$ one also has 
\be
N = \sum_{\bf k} \langle n_{\bf k} \rangle = \sum_{\bf k} \Theta(\mu-\epsilon_{\bf k}) ,\label{Nmu}
\ee 
which is also the $T=0$ limit of \eqref{Nmut}. This implies that $\tilde \mu$ is related to
$\mu$ by equating the two relations \eqref{Nmut} and~\eqref{Nmu}.

\medskip

{\bf Bulk behavior:} Substituting the result \eqref{W0bulk} for the bulk zero temperature Wigner function in \eqref{relationW}
we obtain
\be
W_{\tilde \mu}({\bf x},\p) = \frac{1}{1 + e^{\beta ( \frac{\p^2}{2 m } + V(\bf x) - \tilde \mu)}} \;. \label{WTbulk} 
\ee 
This equation is valid for all ${\bf x},{\bf p}$ in the phase space 
where $\frac{\p^2}{2 m } + V({\bf x}) - \tilde \mu = O(T)$. In the canonical ensemble
$\tilde \mu$ is related to $N$ via \eqref{Nmut}, and using also \eqref{Nmu}, one sees that
the bulk regime corresponds to
scaling $T \sim \tilde \mu$ and $T \sim \mu$. By integrating \eqref{WTbulk} over ${\bf p}$ (respectively
over ${\bf x}$) one obtains the finite temperature bulk density $\rho_N(\x)$ (respectively momentum density
$\bar \rho_N(\p)$ (see e.g. Eqs. (270-273) in \cite{fermions_review}). 

\medskip

{\bf Edge behavior:}  For simplicity let us first focus on the
harmonic oscillator. There it turns out that in the edge regime,
one needs to scale the temperature as $T \sim \mu^{1/3}$
\cite{fermions_review} in the limit of large $\mu$. Consequently one defines a reduced
inverse temperature 
\be\label{def_b_H}
b=\beta \mu^{1/3},
\ee
with $b=O(1)$ in this regime. In addition, in this regime $\tilde \mu \simeq \mu$ 
\cite{fermions_review}, hence we set $\tilde \mu=\mu$ in the following.  
Our starting point is the integral in \eqref{relationW}. For 
$T \sim \mu^{1/3}$, this integral is dominated by the regime where
$\mu'$ is close to $\mu$. In fact by setting
\be
\beta(\mu'-\mu)= - b u,  \label{defu} 
\ee
we see that the Fermi factor in \eqref{relationW} takes the dimensionless form $1/(1 + e^{-b u})$. 
This suggests that the integral will be controlled by values of $u$ which are  of order unity, 
hence 
\be\label{fermi_finiteT}
\mu'-\mu = O(\mu^{-1/3}) \;.
\ee
Therefore, at any point ${\bf x}, {\bf p}$ in phase space close to the 
edge defined in \eqref{edgeeq}, we can use the scaling form \eqref{scal1}-\eqref{scal2}
of $W_N({\bf x},\p)$ inside the integral in \eqref{relationW}. We note that
the only dependence on $\mu'$ of ${\cal W}(a)$ is through the scaling variable
\be
a=a_{\mu'} :=\sqrt{2} \, (\mu')^{1/6} \left(\sqrt{\p^2 + \x^2}- \sqrt{2 \mu'}\right). \label{amu1} 
\ee
Using the scaling form we have
\be
\partial_{\mu'} W_N({\bf x},\p) \simeq {\cal W}'(a_{\mu'}) \partial_{\mu'} a_{\mu'} .\label{der1} 
\ee
We can now expand $a_{\mu'}$ in \eqref{amu1} around $\mu'=\mu$ as
\be
a_{\mu'} = a_{\mu} - \frac{1}{\mu^{1/3}} (\mu'-\mu) .
\ee
Using \eqref{defu} and $b=\beta \mu^{1/3}$ we obtain 
\be
a_{\mu'} = a_{\mu} + u + O( \mu^{-2/3}). \label{der2}
\ee 
Inside the integral in \eqref{relationW} the factor 
\be
\partial_{\mu'} a_{\mu'} = \frac{du}{d \mu'} \partial_u a_{\mu'} = \frac{du}{d \mu'} (1 + O( \mu^{-2/3})).
\ee
We now rewrite the integral in \eqref{relationW} in terms of the $u$ variable using 
\eqref{der2} in the argument of ${\cal W}'$ in the r.h.s. of \eqref{der1}. This gives us
the finite temperature Wigner function, which near the edge, takes the scaling form
\be
W_N({\bf x},\p)  \simeq \frac{{\cal W}_b(a)}{(2 \pi \hbar)^d}  \, , \quad   a = \frac{1}{w_N} \left(\sqrt{\frac{\p^2}{m^2 \omega^2} + \x^2}- r_e\right) \label{scal1T}
\ee
where $w_N$ is given in \eqref{wN1}. The 
finite temperature scaling function, parameterized by $b=\beta \mu^{1/3}$ is 
\bea
&&   
 {\cal W}_b(a) = \int_{-\infty}^{+\infty} \frac{2^{2/3} du}{1+ e^{-b u}} \Ai(2^{2/3} (u+a)) \;. \label{scal2T} 
\eea
It reduces to the $T=0$ scaling form in the limit $b \to \infty$ where the Fermi factor
becomes a Heaviside theta function. The asymptotics of ${\cal W}_b(a)$ can be computed easily. After making the change of variable $2^{2/3}(u+a) = y$, we obtain
 \bea\label{asympt_Wb1} 
 {\cal W}_b(a) = \int_{-\infty}^\infty dy\, \frac{{\rm Ai}(y)}{1 + e^{a\,b}e^{-b\,y 2^{-2/3}}} \;.
\eea
For $a \to -\infty$, this gives 
\bea\label{asympt_Wb2}
{\cal W}_b(a) \to \int_{-\infty}^\infty dy \, {\rm Ai}(y) = 1 \;.
\eea
In contrast, when $a \to + \infty$, we get to leading order
\bea\label{asympt_Wb3}
{\cal W}_b(a) \sim C\, e^{-a\,b} \;, \;
\eea
where the prefactor 
\bea\label{asympt_Wb4}
C = \int_{-\infty}^\infty dy \, {\rm Ai}(y) \, e^{b\,y\,2^{-2/3}} = e^{b^3/12} \;.
\eea

As in the zero temperature case discussed in section \ref{sec:V}, 
we can extend the above finite temperature results to the case 
of arbitrary smooth potentials. As discussed in section \ref{sec:V}
the zero temperature scaling form ${\cal W}(a)$ of the Wigner
function near the edge is identical to that of the harmonic oscillator.
In contrast the scaling variable takes the form 
\bea
a = a_{\mu'} = \frac{1}{e_N(\mu')} \left(\frac{\p^2}{2 m} + V({\bf x}) -  \mu'\right), \label{aV2} 
\eea 
where the energy scale $e_N(\mu')$ is an implicit function of $\mu'$ which
can be obtained from formula \eqref{eN}. It thus depends non-universally on the
shape of the potential $V(\x)$ and on the precise location of $\x_e,\p_e$ 
on the edge surface in the 
phase space. Following the same steps as for the harmonic oscillator case, 
the reduced inverse temperature variable is now
\be
b = \beta \, e_N(\mu).
\ee
Defining $u$ as in \eqref{defu} we can again expand $a_{\mu'}$ in \eqref{aV2} 
around $\mu'=\mu$ as
\be
a_{\mu'} \simeq a_{\mu} - \left[ \frac{1}{e_N(\mu)} + a_\mu \partial_\mu \ln e_N(\mu) \right] (\mu'-\mu), 
\ee
which leads exactly to Eq. \eqref{der2} to leading order in $\mu$. The rest of
the argument simply goes through and we find that, for arbitrary smooth potentials,
the finite temperature Wigner function near the edge takes the form
\be
W_N({\bf x},\p)  \simeq \frac{{\cal W}_b(a)}{(2 \pi \hbar)^d}  \, , \quad   a = 
\frac{1}{e_N} \left(\frac{\p^2}{2 m} + V({\bf x}) -  \mu\right),
\ee 
with exactly the same scaling function ${\cal W}_b(a)$ given in~\eqref{scal1T}.

\section{Conclusion}

In this paper we have studied the Wigner function $W_N(\x,\p)$ for $N$ non interacting fermions in 
a confining trap in $d$ dimensions. At zero temperature and large $N$, we have shown that there
are two main regimes for $W_N(\x,\p)$ in the phase space. A bulk regime where
the Wigner function is flat over a finite support and vanishes outside.
The edge of this support in the phase space is given by $\frac{\p^2}{2 m} + V(\x)=\mu$,
where $\mu$ is the Fermi energy. Around this edge $W_N(x,p)$, appropriately
centered and scaled, is described by a scaling function ${\cal W}$.
We have shown that this scaling function is universal, i.e. the same for 
a large class of confining smooth potentials, and, strikingly, is independent of the space
dimension $d$. We then extended these results to finite temperature 
and found a one-parameter edge scaling function ${\cal W}_b$
where $b$ is the scaled temperature. The finite temperature scaling
function is also universal and independent of the space dimension $d$.
In both the bulk and the edge regimes, the scaling functions are non-negative everywhere.

In addition to these universal features (bulk and edge) there appears to be
anomalous small scale regimes around special points in the phase space where 
the Wigner function is rapidly varying. For the harmonic oscillator in $d=1$ we have
analyzed such an anomalous regime \cite{berry1} near $x=p=0$ in detail. Indeed it is well-known,
already at the single particle level,
that the Wigner function is not guaranteed to be non-negative for arbitrary potentials
$V(x)$. However, for the scaling regimes at large $N$ (bulk and edge), our results show
that the Wigner function remains positive, for a large class of smooth potentials.

Finally, in this paper, we have focused on the standard Wigner function $W_N(\x,\p)$ of an $N$-body system, 
as defined in Eq. (\ref{wig1def}). This can be interpreted as the one-point probability density function in the phase space (in the semi-classical sense). Naturally, one can also investigate higher order correlation functions in the phase space. At $T=0$, 
this can be naturally done by introducing a generalized Wigner function \cite{wigner}
\begin{eqnarray}\label{gen_Wigner}
&&W_{N}^{(N)}({\bf x}_1,{\bf p}_1,\cdots \,, {\bf x}_N,{\bf p}_N) =\nonumber \\
&&\frac{1}{(2\pi\hbar)^{N\,d}}\int d{\bf y}_1\cdots d{\bf y}_N \ e^{\frac{i}{\hbar}\sum_{i=1}^N {\bf p}_i\cdot {\bf y}_i}\nonumber \\ && \Psi^*_0({\bf x}_1+\frac{{\bf y}_1}{2},\cdots\,, {\bf x}_N+\frac{{\bf y}_N}{2})\Psi_0({\bf x}_1-\frac{{\bf y}_1}{2},\cdots\,, {\bf x}_N-\frac{{\bf y}_N}{2}) \;,\nonumber \\
\end{eqnarray} 
where $\Psi_0$ is the ground state many-body wave function. From this generalized Wigner function, one can construct successively $n$-point correlation functions by integrating out $N-n$ phase space coordinates as follows   
\begin{eqnarray}\label{def_correlation}
&&C^{(N)}_n({\bf x}_1,{\bf p}_1,\cdots, {\bf x}_n,{\bf p}_n) = \nonumber  \frac{N!}{(N-n)!}\times \\
&&\int d{\bf x}_{n+1}d{\bf p}_{n+1}\cdots d{\bf x}_{N}d{\bf p}_{N} W_{N}^{(N)}({\bf x}_1,{\bf p}_1,\cdots \,,  {\bf x}_N,{\bf p}_N).\nonumber\\
\end{eqnarray}
For instance, for $n=1$, Eq. (\ref{def_correlation}) reduces precisely to the standard Wigner function $W_N(\x,\p)$ defined in Eq.~(\ref{wig1def}), i.e. $C_1^{(N)}(\x,\p) = W_N(\x,\p)$. Investigations of these higher order correlation functions with $n > 1$, both at $T=0$ and $T>0$ would be interesting and will be studied in a future publication \cite{unpublished}.

\acknowledgments

We thank Paul Wiegmann for stimulating discussions about Ref. \cite{Wiegman}. 
This research was supported by ANR grant ANR-17-CE30-0027-01 RaMaTraF.

\appendix

\begin{widetext} 

\section{Derivation of equation \eqref{Kmu2}} \label{app:A} 

We start from the expression of the ground state for $N$ noninteracting fermions as 
a Slater determinant
\be
\Psi_0(\x_1, \x_2,\cdots, \x_N) = \frac{1}{\sqrt{N!}} \det_{1 \leq i,j, \leq N} \psi_{{\bf k}_j}(\x_i)
\ee
We now evaluate the integral on the r.h.s. of Eq. \eqref{Kmu2} as
\bea
&& N \int_{-\infty}^{+\infty} d\x_2 \ldots d\x_N  \Psi_0^*(\x, \x_2,\cdots, \x_N)  \Psi_0(\x', \x_2,\cdots \x_N)  \\
&& = \frac{1}{(N-1)!} \sum_{\sigma,\tau \in {\cal S}_{N}} (-1)^{\sigma \tau} 
\psi^*_{{\bf k}_{\sigma(1)}}(\x) \psi_{{\bf k}_{\tau(1)}}(\x')
 \int_{-\infty}^{+\infty} d\x_2 \ldots d\x_N \prod_{i=2}^N 
 \psi^*_{{\bf k}_{\sigma(i)}}(\x_i) \psi_{{\bf k}_{\tau(i)}}(\x_i) \\
&& = \frac{1}{(N-1)!}\sum_{\sigma\in {\cal S}_{N}} 
\psi^*_{{\bf k}_{\sigma(1)}}(\x) \psi_{{\bf k}_{\sigma(1)}}(\x')=\sum_{i=1}^N \psi^*_{{\bf k}_{i}}(\x) \psi_{{\bf k}_{i}}(\x') = K_\mu(\x,\x') 
\eea
where  ${\cal S}_{N}$ denotes the group of permutations over $N$ elements. In the middle line each integral over $x_i$, $i=2,\cdots,N$ constrains the permutations in the double sum to be the same, {\em i.e.} $\sigma(i)=\tau(i)$, for each $i$ between $2$ and $N$,  from orthonormality of the single particle eigenfunctions. However this also constrains $\sigma_(1)=\tau(1)$ and we thus have $\sigma=\tau$. The sum in the last line is obtained
by setting $\sigma(1)=i$ and summing over $\sigma$. Since we are dealing with the
ground state the ${\bf k}_i$ correspond to the $N$ lowest eigenstates, recovering
\eqref{Kmu}. 

\section{Edge kernel and Wigner function}\label{appb}

In \cite{fermions_review} it was shown that, for a $d$-dimensional harmonic oscillator at $T=0$, the
kernel near a point on the edge in position space $\x_{em}=\bf{r}_e$ takes the scaling form
\be
K_\mu(\x,\x') \simeq \frac{1}{w_N^d} {\cal K}^{\rm edge}_d\left(\frac{\x - {\bf r}_e}{w_N} , \frac{\x' - {\bf r}_e}{w_N}\right) \label{Kscal1}
\ee
where the scaling function ${\cal K}^{\rm edge}_d({\bf a},{\bf b})$ is given by
\be
{\cal K}^{\rm edge}_d\left({\bf a},{\bf b}\right) = \int \frac{d^d q}{(2 \pi)^d} e^{- i {\bf q} \cdot ({\bf a}-{\bf b})}
{\rm Ai}_1\left(2^{2/3}( q^2 + \frac{a_n + b_n}{2})\right) \label{Kscal2}
\ee
where $a_n= {\bf a}\cdot{\bf r}_e/r_e$ and $b_n={\bf b}\cdot{\bf r}_e/r_e$. In Eq. (\ref{Kscal2}) we have  ${\rm Ai}_1(x) = \int_x^\infty du{\rm Ai}(u)$. Setting $\x \to \x - \frac{\bf y}{2}$
and $\x' \to \x' + \frac{\bf y}{2}$, gives, after rescaling ${\bf q} \to {\bf q} w_N$,
\bea
K_\mu\left(\x - \frac{\bf y}{2},\x + \frac{\bf y}{2}\right) \simeq 
 \int \frac{d^d q}{(2 \pi)^d} e^{- i {\bf q} \cdot {\bf y}} 
{\rm Ai}_1\left(2^{2/3}\left( w_N^2 q^2 + \frac{({\bf x}- {\bf r}_e) \cdot{\bf r}_e}{w_N r_e}\right)\right).
\eea
Fourier transforming w.r.t. ${\bf y}$ we obtain
\bea
W_N(\x,\p) \simeq \frac{1}{(2 \pi \hbar)^d} {\rm Ai}_1\left(2^{2/3}( \frac{w_N^2}{\hbar^2} {\bf p}^2 + \frac{({\bf x}- {\bf r}_e)\cdot{\bf r}_e}{w_N r_e})\right) = \frac{1}{(2 \pi \hbar)^d}{\cal W}\left(\frac{w_N^2}{\hbar^2} {\bf p}^2 + \frac{({\bf x}- {\bf r}_e)\cdot{\bf r}_e}{w_N r_e}\right)
\eea 
from Eq. (\ref{scal2}). Furthermore we identify the argument which appears in the 
scaling function as $2^{2/3} a$, where $a$ has been expanded around the point ${\bf x}={\bf r}_e$, ${\bf p}=0$
of the edge surface in the phase space. Indeed, to lowest order in $({\bf x}- {\bf r}_e)$ and $\p^2$ one has
\bea
&& a = \frac{1}{w_N} (\sqrt{\frac{\p^2}{m^2 \omega^2} + \x^2}- r_e) 
 \simeq  \frac{({\bf x}- {\bf r}_e)\cdot{\bf r}_e}{w_N r_e} + \frac{w_N^2}{\hbar^2} \p^2 
\eea
using the relation $w_N^3 r_e/\hbar^2 = 1/(2 m^2 \omega^2)$ valid for the harmonic oscillator.
This provides an alternative derivation of $W_N({\bf x}, {\bf p})$ which is valid near the
special point ${\bf x}_e={\bf r}_e$, ${\bf p}_e=0$ on the edge surface in the phase space. 
However for the harmonic oscillator, since $W_N(\x,\p)$ is isotropic in the phase space
(in dimensionless units)  it clearly  suffices to establish the result \eqref{scal1} for
any point on the edge surface. This result is thus fully consistent with our
derivation of the scaling behavior of the Wigner function for the harmonic oscillator.

Furthermore, using the fact that the scaling form \eqref{Kscal1}, \eqref{Kscal2} is universal
for a broad class of smooth potentials, one can similarly show that the
zero temperature Wigner scaling function is also universal around the special point ${\bf x}={\bf r}_e$, 
${\bf p}_e=0$, and matches with the formula for $e_N$, given in Eq. (\ref{eN_intro}) at this special point only.
However, since for a general potential the isotropy of the Wigner function in phase space
no longer holds, this method does not allow one to obtain the scaling form at a generic point
on the edge surface. However, the method used in the text does not rely on isotropy 
and is valid anywhere on the edge surface. 


\section{Short time expansion and universality for a class of smooth potentials}\label{App:short_time}

In this Appendix, for simplicity, we use units such that $m=\hbar=1$ and restore the units in the text. Consider
a generic point in phase space ($\x_e,\p_e$) on the edge surface defined by
\be
 \frac{\p_e^2}{2} + V({\bf x}_e) = \mu. \label{edgeeq2}
\ee 
We demonstrate (the result given in the main text) that for $\x,\p$ near such a point, and for  the scaling variable $a=O(1)$ with $a$ defined as
\be
a = \frac{1}{e_N} (\frac{\p^2}{2} + V({\bf x}) -  \mu)  \quad , \quad e_N = \frac{1}{2^{1/3}} \left((\p_e \cdot  \nabla)^2 V({\bf x}_e)  + |\nabla V({\bf x}_e)|^2 \right)^{1/3}, \label{ae_app} 
\ee
the Wigner function takes the form
\be
W_N({\bf x},\p)  \simeq \frac{{\cal W}(a)}{(2 \pi \hbar)^d} .   \label{scal1V2}
\ee

This statement is valid in the limit of large $\mu$,
which can be studied using the short time expansion of the Euclidean propagator, extending
the calculation performed in the Appendix A of \cite{fermions_review} (see below). 
It is useful to anticipate the main idea of the proof. First, at a generic point \eqref{edgeeq2} 
one has $|\p_e| \sim \mu^{1/2}$ and $V(\x_e) \sim \mu$. The two terms in the 
energy scale $e_N$ in \eqref{ae_app} are thus both of the same order, with $e_N \sim 
|\nabla V({\bf x}_e)|^{2/3}$. The typical time scale $t=t_N$ which will control the final integral over $t$ (see below in Eq. (\ref{starting})) is $t_N \sim 1/e_N$, which, in the particular case $\p_e=0$, also agrees with the result given by Eq. (282) in \cite{fermions_review}. For a potential $V(\x) \sim |\x|^p$ at large $|\x|$ the estimate is 
$e_N \sim (\mu/x_e)^{2/3} \sim \mu^{2 (p-1)/(3 p)}$ (with $x_e= |\x_e|$) and $t_N \sim (x_e/\mu)^{2/3} \sim \mu^{-2 (p-1)/(3 p)}$,
consistent for $p=2$ with $t_N \sim \mu^{-1/3}$ obtained for the harmonic oscillator in the text. 
We will justify these statements below, but it is useful to
keep them in mind for estimating the various terms. 

Before performing the short time expansion, let us first derive some useful exact representations for
the Wigner function. We use the relation \eqref{Wdirect} between the Wigner function and the Euclidean propagator
\bea
&&  W_N({\bf x},\p) = \frac{1}{(2\pi)^d}  
 \int_{C} \frac{dt}{2 \pi i t} e^{\mu t} \int_{-\infty}^{+\infty} d\y e^{i \p \cdot {\bf y}}\, G(\x+ \frac{\y}{2} ,\x - \frac{\y}{2}, t). \label{starting} 
\eea
From the Appendix A of \cite{fermions_review}
we can write, as an exact starting point, the following representation (using the symmetry of the Euclidean propagator) 
\bea \label{exact1} 
G(\x- \frac{\y}{2} ,\x + \frac{\y}{2}, t) = \frac{1}{(2 \pi t)^{d/2}} \exp[ - \frac{{\bf y}^2}{2 t} ]
 \left\langle \exp\left(- t \int_0^1 du\  V( {\bf x} + {\bf y} (u-\frac{1}{2}) + \sqrt{t}{\sf B}_u)\right) \right\rangle_{{\sf B}}
\eea 
where $\left\langle \dots  \right\rangle_{{\sf B}}$ denotes an average over the
$d$-dimensional Brownian bridge ${\sf B}_u=\{B_{iu}\}_{i=1}^d$ on the interval $[0,1]$, i.e. a Gaussian process with mean zero and correlation function
\begin{equation}
\left\langle B_{iu} B_{ju'}\right\rangle_{{\sf B}} = \delta_{ij} \ g(u,u') \quad , \quad g(u,u') = \min(u,u') - u u'
\end{equation}
hence with ${\sf B}_0={\sf B}_1=0$. By first performing a cumulant expansion 
and then  expanding in $\sqrt{t}{\sf B}_u$,  we generate the short-time expansion of Eq. (\ref{exact1}). 

We now substitute \eqref{exact1} into \eqref{starting}. We note that, in the absence of a potential (or if we neglect
the ${\bf y}$ dependence in the potential term) we
have a Gaussian integral over ${\bf y}$ with a saddle point at ${\bf y}= i t {\bf p}$.
This suggests that  it is natural to make the change of integration variable
\be
{\bf y}= i t {\bf p} + \tilde {\bf y} \sqrt{t} 
\ee 
and rewrite
\bea
&&  W_N({\bf x},\p) = \frac{1}{(2\pi)^{d}}  
 \int_{C} \frac{dt}{2 \pi i t} e^{(\mu - \frac{{\bf p}^2}{2} - V({\bf x}) ) t + S({\bf x},\p,t)} \label{WS} 
\eea
where
\bea
&& S({\bf x},{\bf p},t) = \ln
 \left\langle \exp\left(- t \int_0^1 du\  [ V( {\bf x} + ( i t {\bf p} + \tilde {\bf y} \sqrt{t}) 
 (u-\frac{1}{2}) + \sqrt{t}{\sf B}_u) - V({\bf x}) ] \right)  \right\rangle_{{\sf B},\tilde {\bf y} } ,\label{defS} 
\eea
where the "average" over $\tilde {\bf y}$ is over a unit Gaussian random variable, uncorrelated 
with ${\sf B}$,
i.e.
\bea
 \left\langle \dots  \right\rangle_{{\sf B},\tilde {\bf y} } 
 = \int_{-\infty}^{+\infty} \frac{d \tilde \y}{(2 \pi)^{d/2}} 
 \exp[ - \frac{\tilde {\bf y}^2}{2} ]   \left\langle \dots  \right\rangle_{{\sf B}} .
\eea 
%
Since the averaging measure is even in $\tilde {\bf y}$ and even in ${\sf B}_u$ it is clear that 
$S({\bf x},{\bf p},t)$ starts at $O(t^2)$. Hence the form \eqref{WS} is quite convenient
to study the short time expansion. The leading term, $O(t)$ in the exponential, is obtained by 
setting $S({\bf x},{\bf p},t)$ to zero, which recovers the 
result Eq. (\ref{W0bulk}) of the text for the Wigner function in the bulk. 

%
%
%
%
%
%
%

Since $\tilde \y$, ${\sf B}$ and $\p$ do not depend on $t$, the $t$ dependence of 
$S({\bf x},{\bf p},t)$ in \eqref{defS} is explicit, and its expansion in powers of $t$ at small time is straightforward, although tedious. It is done by a gradient expansion of the argument of $V$ around $\x \simeq \x_e$. 
In doing so we also need to check that this gradient expansion is consistent 
with the expansion at large $\mu$, i.e
that in the large $\mu$ limit all terms in the argument of $V$ are small compared to $\x_e$. This is clearly the
case for the terms proportional to $\tilde {\bf y}$ and ${\sf B}_u$, for which the gradient
expansion is an expansion in the parameter $t_N^{1/2}/x_e \sim \mu^{-1/3} x_e^{-2/3} \ll 1$, using
our above anticipated estimate for $t_N$. 
For the term $i t {\bf p}$, the gradient expansion parameter is $t_N p_e/x_e \sim 
\mu^{-1/6} x_e^{-1/3} \ll 1$. 

We now calculate $S({\bf x},{\bf p},t)$ up to $O(t^3)$.
For this it is easy to see that we need only the first two cumulants in \eqref{defS}.  The first cumulant is
\bea
S_1({\bf x},{\bf p},t) &=& - t \int_0^1 du  [ 
 \left\langle  V( {\bf x} + ( i t {\bf p} + \tilde {\bf y} \sqrt{t}) 
 (u-\frac{1}{2}) + \sqrt{t}{\sf B}_u))  \right\rangle_{{\sf B},\tilde {\bf y} } - V({\bf x}) ] \\
 & =& - t^2 \left[ 
 \left(  \frac{1}{24} 
 \langle \tilde y_j \tilde y_k \rangle_{\tilde {\bf y} } 
 + \frac{1}{2} \int_0^1 du \langle B_{ju} B_{ku}   \rangle_{{\sf B}} \right)  \nabla_j \nabla_k   V({\bf x}) \right] \\
&& - t^3 \bigg[ - \frac{1}{8} \int_0^1 du (1-2 u)^2 ({\bf p} \cdot \nabla)^2 V(\x) 
 + \bigg( \frac{1}{1920}  
\langle \tilde y_j \tilde y_k \tilde y_\ell \tilde y_n\rangle_{\tilde {\bf y} } \nn
\\
&& + \frac{1}{16}  \langle \tilde y_j \tilde y_k \rangle_{\tilde {\bf y} }
\int_0^1 du (1- 2 u)^2 \langle {B}_{\ell u} {B}_{n u}   \rangle_{{\sf B}} 
 + \frac{1}{24} 
\int_0^1 du \langle {B}_{ju} {B}_{ku} {B}_{\ell u} {B}_{n u}   \rangle_{{\sf B}} \bigg)
\nabla_j \nabla_k \nabla_\ell  \nabla_n V({\bf x}) \bigg] + O(t^4) \nn 
\eea.
Note that all terms proportional to an odd number of gradients vanish due to the
symmetry $u \to 1-u$ in the integral over $u$. Performing all the averages we obtain 
\bea
S_1({\bf x},{\bf p},t) =  \label{S1exp}
 &-& t^2   (  \frac{1}{24}  + \frac{1}{12} )  \nabla^2 V({\bf x}) \\
&-& t^3 \left[ - \frac{1}{24} ({\bf p} \cdot \nabla)^2 V(\x)  +
(\frac{1}{640} + \frac{1}{480} + \frac{1}{240} )  \nabla^2 \nabla^2 V({\bf x}) \right], \nn  
\eea 
where we have purposely indicated the contribution of each term, and checked that
the last term is identical to the result in Eqs. (246-248) in \cite{fermions_review}. The second cumulant 
expanded to $O(t^3)$ reads
\bea
S_2({\bf x},{\bf p},t) &=& \frac{t^2}{2} \int_0^1 du \int_0^1 du' \label{S2exp} \\
&& \!\!\!\!\!\!\!\!\!\!\!\!\!\!\!\! \!\!\!\!\!\!\!\!\!\!\!\!\!\!\!\! \times  
 \left\langle  [V( {\bf x} + ( i t {\bf p} + \tilde {\bf y} \sqrt{t}) 
 (u-\frac{1}{2}) + \sqrt{t}{\sf B}_u)) - V({\bf x}) ]   [ 
 V( {\bf x} + ( i t {\bf p} + \tilde {\bf y} \sqrt{t}) 
 (u'-\frac{1}{2}) + \sqrt{t}{\sf B}_{u'})) - V({\bf x}) ]   \right\rangle^c_{{\sf B},\tilde {\bf y} } \\
 &= &\frac{t^3}{2} \int_0^1 du \int_0^1 du'  [ (u-\frac{1}{2})  (u'-\frac{1}{2})  
\tilde y_j \tilde y_k 
+  B_{ju}  B_{ku'}]  \nabla_j V({\bf x}) \nabla_k V({\bf x}) + O(t^4) \\
&= & \frac{t^3}{24} |\nabla V({\bf x})|^2 + O(t^4) .
\eea 
The sum $S({\bf x},{\bf p},t) = S_1({\bf x},{\bf p},t) + S_2({\bf x},{\bf p},t) + O(t^4)$
together with \eqref{S1exp} and \eqref{S2exp} provides the exact short time expansion
up to $O(t^3)$ of $S({\bf x},{\bf p},t)$ which enters the formula \eqref{WS} for the 
Wigner function (it gives in fact the short time expansion
up to $O(t^3)$ of the Fourier transform of the Euclidean propagator $G(\x- \frac{\y}{2} ,\x + \frac{\y}{2}, t)$).  

Now remember that our goal is instead the large $\mu$ expansion of the Wigner function.
If one can show that (i) all terms in $S_1$ except  the term $\frac{t^3}{2} ({\bf p} \cdot \nabla)^2 V(\x)$
are irrelevant in the edge regime (ii) all terms $O(t^4)$ or higher are also irrelevant, then
we see that 
\bea
&&  W_N({\bf x},\p) \simeq \frac{1}{(2\pi)^{d}}  
 \int_{C} \frac{dt}{2 \pi i t} e^{(\mu - \frac{{\bf p}^2}{2} - V({\bf x}) ) t + 
 \frac{t^3}{24} ( |\nabla V({\bf x}_e)|^2 
 +  ({\bf p}_e \cdot \nabla)^2 V(\x_e) ) },
 \label{WSn} 
\eea
where in the cubic term it is consistent to replace $\x$ by $\x_e$. Performing the change of variable
$t= 2^{2/3} \tau/e_N$, with $e_N$ given by \eqref{ae_app}, we obtain exactly the integral representation
\eqref{rep} of the function ${\cal W}(a)$, hence demonstrating the  result \eqref{scal1V2}
with the scaling variable $a$ defined in \eqref{ae_app}. Furthermore this confirms that
$t_N \sim 1/e_N$ is the time scale which dominates the integral in the edge regime
as anticipated above. 

Estimating the term $O(t^2)$ in \eqref{S1exp} to be of order 
$t_N^2 V(\x_e)/x_e^2 \sim \mu^{-1/3} x_e^{-2/3} \ll 1$ we see that it is indeed
negligible, as was already the case for $\p_e={\bf 0}$  in \cite{fermions_review}.
Inside the $t^3$ term in \eqref{S1exp} we see that the second term is
smaller than the first by a factor $\frac{1}{p_e x_e^2} \sim \mu^{-1/2} x_e^{-2} \ll 1$. 
The examination of terms $O(t^4)$ and higher is very tedious and can be performed along
the lines of the Appendix A in \cite{fermions_review}. We will not reproduce this
analysis here.

In summary, the above shows that for a large class of smooth potentials, 
at a generic point of the edge surface in phase space, the universal edge
form of the Wigner function holds. The analysis
is a rather simple extension of the one in \cite{fermions_review} (simple
in the sense that the characteristic scales are not changed, apart from some
pre-factors). Note that a necessary condition is that $e_N$ does not vanish, 
which is true at a generic point, but could fail in some exceptional cases,
for instance if $\p_e$ and $\nabla V(\x_e)$ vanish simultaneously. 

\section{Wigner function close to the center $(x=0, p=0)$}\label{Appendix:anomalous}

In this Appendix, we show that for the $1d$ harmonic oscillator at $T=0$ the Wigner function near $(x=0, p=0)$ in the
phase space has an anomalous behavior. Indeed, for $r^2 = x^2 + p^2 = O(1/N)$, we will show that $W_N(x,p)$ has the following behavior 
\beq\label{anomalous_app}
W_N(x,p) \sim \frac{1}{2\pi} - (-1)^N F(\sqrt{N(x^2+p^2)}) \;, \; F(z) = \frac{1}{2\pi} J_0( 2\sqrt{2} z) \;,
\eeq
where $J_\nu(x)$ is the Bessel function of index $\nu$. Our starting point is the exact generating function in Eq.~(\ref{exactGF}). 
\begin{figure}[ht]
\includegraphics[width = 0.45 \linewidth]{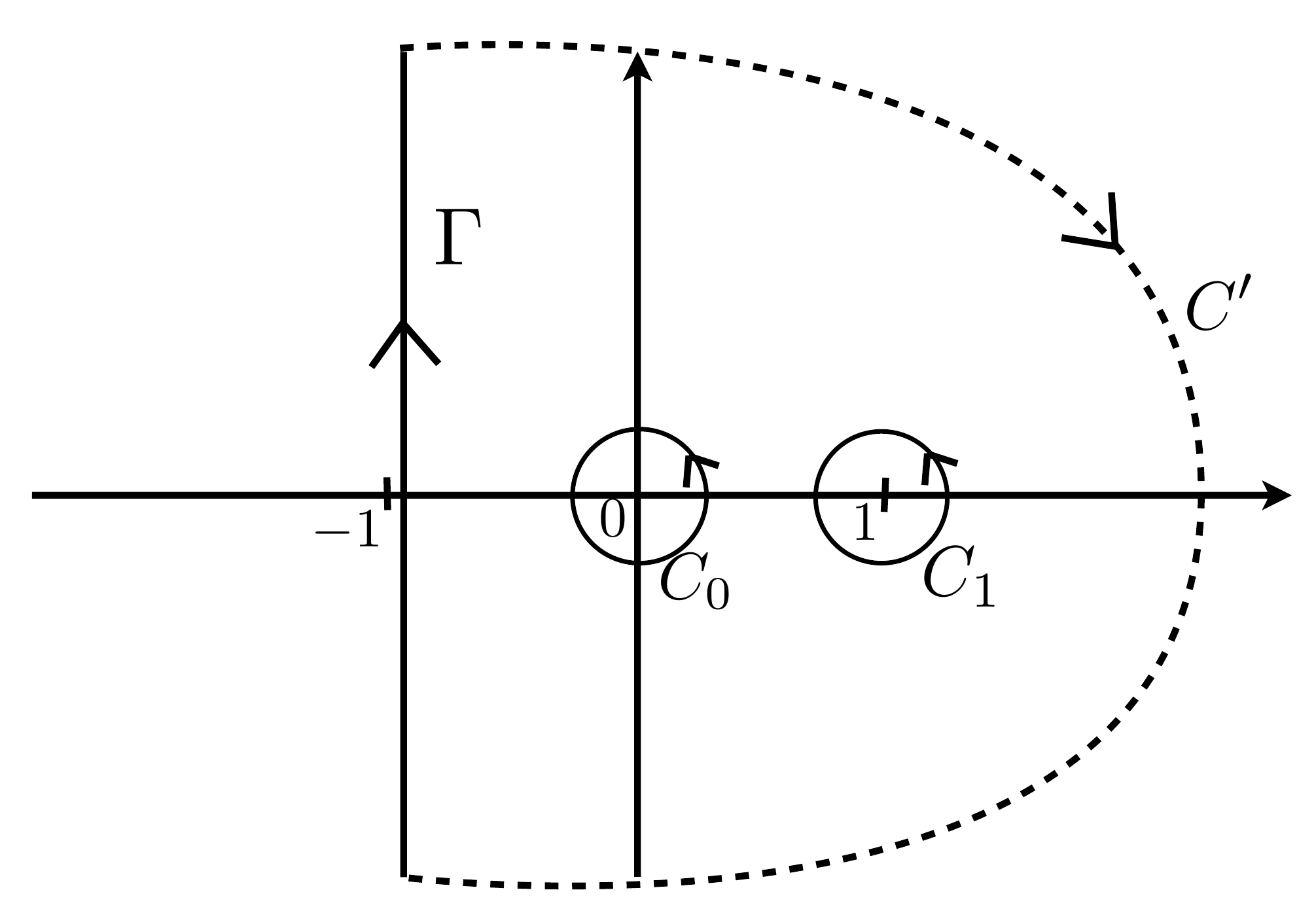}
\caption{The contours in the complex $z$-plane used to evaluate the integral in Eq. (\ref{Cauchy}) using the decomposition in Eq. (\ref{three_contours}).}\label{Fig:contour}
\end{figure}
Formally inverting this generating function using Cauchy's formula, we find
\beq\label{Cauchy}
W_N(x,p) = \frac{1}{2 \pi i} \int_{C_0} \frac{dz}{z^{N}} \frac{1}{\pi(1-z^2)}\,e^{-\frac{1-z}{1+z}r^2} \;,
\eeq 
where $C_0$ is the contour around the origin in the complex $z$-plane, as shown in Fig. \ref{Fig:contour}. For $z$ such that $\Re(z) > - 1$, the integrand in Eq. (\ref{Cauchy}) has a simple pole at $z = 1$ and an $N$-th order pole at $z=0$ and is analytic elsewhere. We can thus replace the contour integral on $C_0$ by three other contour integrals as follows (see Fig. \ref{Fig:contour})
\beq\label{three_contours}
\int_{C_0} = -\int_{C_1} - \int_{C'} - \int_{\Gamma} 
\eeq
 where the contours $C_1, C'$ and $\Gamma$ are shown in Fig. \ref{Fig:contour}. We will eventually deform the contours $C'$ and $\Gamma$ such that $\Gamma$ is a straight vertical line passing infinitesimally close to the right of $z=-1$ and $C'$ will be eventually sent to infinity. Evaluating the simple pole around $z=1$ gives
\beq\label{C1}
-\int_{C_1} = \frac{1}{2 \pi} \;.
\eeq
The contribution from the contour integral $C'$ is exponentially small for large $N$ when the contour $C'$ is sent to infinity. It remains to evaluate the contour integral over $\Gamma$. It is clear that the leading contribution to this integral over $\Gamma$ comes from the vicinity of $z=-1$. Hence, it is natural to make the change of variable $z = - 1 + z'$ and note
that the dominant contribution comes from the regime close to $z'=0$, i.e. $z'= O(1/N)$ for large $N$. Therefore we set $z' = t/N$ and expand the integrand for large $N$. To leading order, we obtain
\beq\label{contour_gamma}
W_N(x,p) \sim \frac{1}{2 \pi} - \frac{(-1)^N}{2 \pi} \frac{1}{2\pi i} \int_{\Gamma} \frac{dt}{t} e^{N\,t - \frac{2}{t}r^2} \;.
\eeq
Using the integral representation of the Bessel function $J_0$
\beq\label{J0}
\frac{1}{2\pi i} \int \frac{dt}{t} \,e^{Nt - \frac{y}{t}} = J_0(2 \sqrt{y\,N}) \;,
\eeq
Eq. (\ref{contour_gamma}) immediately gives the result in Eq. (\ref{anomalous_app}).

\end{widetext}


{}

\end{document}